\documentclass[structabstract]{aa}  
\usepackage{graphicx}
\usepackage{txfonts}

\begin{document}

\title{PMAS optical integral field spectroscopy of luminous infrared
  galaxies.}
\subtitle{
 I.-- The atlas\thanks{Based on observations collected at
    the German-Spanish Astronomical Center, Calar Alto, jointly
    operated by the Max-Planck-Institut f\"ur Astronomie Heidelberg and
    the  Instituto de Astrof\'{\i}sica de Andaluc\'{\i}a (CSIC).}}

\author{Almudena Alonso-Herrero\inst{1}, 
Macarena Garc\'{\i}a-Mar\'{\i}n\inst{2},
Ana Monreal-Ibero\inst{3},
  Luis Colina\inst{1}, Santiago Arribas\inst{1}, 
Julia Alfonso-Garz\'on\inst{4}, 
Alvaro Labiano\inst{1}}

\institute{Departamento de Astrof\'{\i}sica Molecular 
e Infrarroja, Instituto de Estructura de la Materia, CSIC, Serrano
121, 
E-28006 Madrid, Spain\\
\email{aalonso@damir.iem.csic.es}
\and
I. Physikalisches Institut, Universit\"at zu K\"oln, Z\"ulpicher
Strasse 77, 50937 K\"oln, Germany
\and
European Organisation for Astronomical Research in the Southern
Hemisphere, Karl-Schwarzschild-Strasse 2 D-85748 
Garching bei M\"unchen, Germany
\and
Centro de Astrobiolog\'{\i}a  (CSIC-INTA), P.O. Box
78, E-28691 Villanueva de la Ca\~nada, Madrid, Spain}

\date{Preprint}

\abstract
{Luminous and ultraluminous infrared galaxies (LIRGs and ULIRGs) are
   key cosmological
  classes since they account for most of the  
co-moving star formation rate density at $z\sim 1-2$. It is then
important to have detailed studies of local samples of their 
counterparts for understanding the internal
and dynamical processes taking place at high-$z$. }
{To characterize the two-dimensional morphological, excitation and
  kinematic properties of LIRGs and ULIRGs we are carrying out an
  optical integral field spectroscopy (IFS) survey of local
  ($z<0.26$) samples.}
{In this paper we present  
optical (3800-7200\AA) IFS with the Potsdam multi-aperture 
spectrophotometer (PMAS) of the northern hemisphere portion of a 
volume-limited ($2750-5200\,{\rm
  km\,s}^{-1}$) sample of 11 LIRGs.
The PMAS IFS observations 
typically cover the central  $\sim 5\,$kpc and are complemented with our own
existing {\it HST}/NICMOS images.}
{For most LIRGs in our sample, 
 the peaks of the continuum and gas (e.g., H$\alpha$, [N\,{\sc
   ii}]$\lambda$6584) emissions coincide, unlike 
what is observed in local, strongly interacting ULIRGs. The 
only exceptions  are galaxies with
circumnuclear rings of star formation where the most luminous
H$\alpha$ emitting regions are found in the rings rather
than in the nuclei of the galaxies, and the displacements are well
understood in terms of differences in the 
stellar populations. A large fraction of the nuclei of these LIRGs 
are classified as LINER  and intermediate LINER/HII,  or
 composite objects, which is a combination of
  starformation and AGN activity. 
The  excitation conditions of the integrated emission  depend on the
relative contributions of H\,{\sc ii} regions and the diffuse emission
to the line emission over the PMAS FoV. Galaxies dominated by
high surface-brightness 
H\,{\sc ii} regions show integrated H\,{\sc ii}-like excitation. 
A few galaxies show  slightly larger integrated [N\,{\sc
  ii}]$\lambda$6584/H$\alpha$ and 
[S\,{\sc ii}]$\lambda$6717,6731/H$\alpha$ line ratios than the nuclear
ones, probably because of more contribution from the diffuse emission.  
The H$\alpha$ velocity fields over the central
few kpc are generally consistent, at least to first order, with
rotational motions.
The velocity fields of most LIRGs are similar to those of 
disk galaxies, in contrast to the highly perturbed 
fields of most local,
strongly interacting ULIRGs. The peak of the H$\alpha$ velocity
dispersion coincides with the position of the nucleus and 
is likely to be tracing mass. All these results are similar to the
properties of  
$z\sim 1$ LIRGs, and  they highlight the importance of detailed studies of
flux-limited samples of local LIRGs.
}
{}


\authorrunning{Alonso-Herrero et al.}
\titlerunning{PMAS integral field spectroscopy of local LIRGs}
\maketitle

\section{Introduction}
In recent years deep cosmological surveys have been extremely successful
in identifying large samples of high-$z$ galaxies using specific 
wavelengths (e.g., UV, optical, infrared, submillimeter) or combinations of
them. Luminous  
infrared galaxies (LIRGs, with infrared $8-1000\,\mu$m 
luminosities $L_{\rm IR}= 10^{11}-10^{12}\,{\rm
  L}_\odot$, see Sanders \& Mirabel 1996) 
and ultraluminous infrared galaxies (ULIRGs, 
with IR luminosities 
$L_{\rm IR}= 10^{12}-10^{13}\,{\rm
  L}_\odot$, see Lonsdale, Farrah, \& Smith 2006) 
are significant cosmological classes.
These IR-selected galaxies are the main contributors to the co-moving
star formation  
rate density of the universe at $z>1$ 
(Elbaz et al. 2002; Le Floc'h et al. 2005; P\'erez-Gonz\'alez et al. 2005; 
Caputi et al. 2007).

Because these high-$z$ samples are so numerous, most efforts concentrate
on characterizing their integrated properties, such
as   stellar 
masses, star formation rates, average ages, and metallicities. 
However, to understand fully how galaxies formed and evolved, one needs
spatially resolved information to study their kinematics,  stellar
populations, the star, gas, and dust distributions, as well as the gas 
excitation conditions. At high-$z$ this has been done for 
optically/UV selected galaxies using integral field spectroscopy
(IFS). These works find
 clumpy H$\alpha$ morphologies with relatively well-ordered 
velocity fields that are consistent with the presence of large disks 
at $z\sim 2$, while other systems are better explained as 
merger candidates (see e.g.,
F\"orster-Schreiber et al. 2006; Genzel et al. 2006, 2008; Wright et 
al. 2009). A similar result is found at 
intermediate redshifts (Puech et al. 2008; Yan et al. 2008). 

In the local universe
LIRGs and ULIRGs are much less numerous than at high-$z$, 
and a large amount of work has already been done to characterize their
  properties using optical long-slit spectroscopy (e.g., Heckman,
  Armus, \& Miley 1987; Armus, Heckman, \& Miley 1989; Kim et al. 1995,
  1999; Veilleux et al. 1995, 1999; Wu et al. 1998; Heckman et
  al. 2000; Rupke, Veilleux \&
  Sanders 2005; Chen et al. 2009). The majority of optical and near-IR 
IFS works have so far focused 
on small samples of (U)LIRGs or individual galaxies 
(e.g.,
 Arribas, Colina, \& Clements 2001; 
Murphy et al. 2001; 
L\'{\i}pari et al. 2004a,b; 
Colina, Arribas, \& Monreal-Ibero 2005; Monreal-Ibero, Arribas, \&
Colina 2006; Garc\'{\i}a-Mar\'{\i}n et al. 2006; Reunanen,
Tacconi-Garman, \& Ivanov 2007).
The work of Shapiro et al. (2008) 
has recently highlighted the importance of having local templates 
for understanding the
internal and dynamical processes taking place at high-$z$. 
We note, however, that care is needed when comparing 
local IR luminosity matched galaxies with high-$z$ systems. For
instance, at $z\sim 2$ the
  mid-IR spectra of star-forming ULIRGs are more similar to those of
  local starbursts and LIRGs than to those of local ULIRGs (Farrah et
  al. 2008; Rigby et al. 2008; Alonso-Herrero et al. 2009). One
  possible explanation is that in high-$z$ ULIRGs star formation 
is taking place over larger scales, a few kpc, than in local ULIRGs
where most of the infrared emission arises from sub-kpc scale regions
(e.g., Soifer et al. 2000). In contrast  intermediate
redshift ($z\sim 0.7$) LIRGs  appear to be experiencing similar
starburst phases to those of local LIRGs (Marcillac et al. 2006).  
We thus  need to  understand  the spatially-resolved physical processes and properties of local
  (U)LIRGs and compare them with those of distant IR-selected galaxies.

\begin{table*}
\begin{center}
\caption{Log of the PMAS Observations.}
\begin{tabular}{lcccccccc}
\hline
Galaxy  &     Dist & $\log L_{\rm IR}$ & Date       &  Galaxy $t_{\rm int}$ &      Sky $t_{\rm int}$  & Airmass & Seeing & Conditions\\  
(1)     & (2)            & (3)                 & (4)
& (5) & (6) & (7) & (8) & (9)\\
\hline
\hline
NGC~23         &  59.6 & 11.05 & 7 Nov 2005 & $3\times 1000$
&  1000  & 1.06& $1.0-1.2$ & P\\ 
MCG~+12-02-001 & 64.3 & 11.44 & 18 Dec 2006 & $4\times 800$ &
1000 &  1.47 &1.7 & NP\\ 
UGC~1845       & 62.0 & 11.07 & 8 Nov 2005 & $3\times 900$  &
900   &  1.05& $1.4-2.0$ & NP\\ 
NGC~2388       & 57.8 & 11.23: & 7 Nov 2005 & $3\times 1000$
& 1000  &  1.04& $1.3-1.1$ & P\\ 
MCG~+02-20-003 & 67.6 & 11.08: & 17 Dec 2006 & $4\times 800$  & $1000$
               &  1.60 & 1.2 & NP\\ 
IC~860        & 59.1 & 11.17: & 29 May 2006 & $4\times 800$ &
1000 &  1.03& 1.7 & NP\\ 
NGC~5936      &  60.8 & 11.07 & 30 May 2006 & $4\times 800$
&  1000 &  1.17& 1.5 & NP\\ 
NGC~6701      &  56.6 & 11.05 &  30 May 2006 & $1\times400 +
4\times 600$   &  800 &1.09 &1.3 & NP\\ 
NGC~7469      &  65.2 & 11.59 & 8 Nov 2005 & $3\times 900$  &
900 &   1.15& 1.2 & NP\\ 
NGC~7591      &  65.5 & 11.05 & 8 Nov 2005 & $3\times 1000$ &
1000 & 1.32& $1.1-1.2$ & NP\\ 
NGC~7771-E    &   57.1 & 11.34 & 7 Nov 2005  & $2\times 500 +
2\times 1000$ & 1000 &    1.15 & 0.8 & P\\ 
NGC~7771-W    &$\cdots$  &$\cdots$  & 7 Nov 2005   & $3 \times 1000$  &
1000 &   1.06  & 0.8 & P\\ 
\hline
\end{tabular} 
\end{center}

Notes.--- Column~(1): Galaxy. For NGC~7771 we took two different
pointings to cover the approximate central $28\arcsec \times
16\arcsec$ region of the galaxy, 
as explained in the text. Columns~(2) and (3): Distance (in
Mpc) and IR 
luminosity (in L$_\odot$) taken from Sanders et al. (2003)
for the {\it IRAS} RBGS. Column~(4): Date of the
observations. Column~(5): Integration time (in seconds) of the galaxy
observations after discarding
bad data sets. Column~(6): Integration
time (in seconds) of the sky observations. Column~(7):  Airmass at the beginning
of the observation. Column~(8): Typical seeing  conditions (in
arcseconds)  of the observations.  If two values of the seeing are given they 
correspond to the beginning and the end of the
observations. Column~(9): Observing conditions. P: Photometric. NP:
Non-photometric.

\end{table*}

We recently  started an optical IFS survey of a representative  
sample of  approximately 70 nearby ($z\le 0.26$) LIRGs and ULIRGs (see
Arribas et al. 2008). 
The general goal of this survey is to provide a two-dimensional, as
opposed to integrated or nuclear, characterization of  the physical
and dynamical processes taking place in local 
LIRGs and ULIRGs. As discussed in detail by Arribas et al. (2008),
we are conducting this survey using three different optical IFS
instruments
in both the northern and the southern hemispheres. These are: 
VIMOS (Le F\'evre et al. 2003) on the VLT, 
the INTEGRAL+WYFFOS system  (Arribas et al. 1998; Bingham et al. 1994) 
on the William Herschel  Telescope,   and 
the  Potsdam multi-aperture spectrophotometer (PMAS, Roth et al. 2005) 
instrument on the 3.5\,m telescope at 
Calar Alto (Spain). In this paper 
we present an atlas of the IFS observations obtained with 
PMAS for the northern hemisphere portion of the volume-limited sample of LIRGs 
of Alonso-Herrero et al. (2006), which is part of the larger 
survey of Arribas et al. (2008). Additionally, we are observing a
representative sample with near-infrared IFS (Bedregal et al. 2009) 
using SINFONI (Eisenhauer et al. 2003) on the VLT.

The paper is organized as follows. Sect.~2 gives details on the
sample, the observations and the 
data reduction. Sect.~3 presents the analysis of
the PMAS IFS data. Sects.~4, 5, and 6, discuss the general results on the
morphologies of the 
emission lines and continuum, the nuclear and integrated 1D spectra,
and the kinematics of the ionized gas, respectively. The
discussion and summary of this work are given in Sect.~7. 
Throughout this paper we used $H_0=75\,{\rm km \, s}^{-1} \, {\rm  Mpc}^{-1}$.

\section{Observations}

\subsection{The Sample}
We observed the majority of the northern hemisphere 
galaxies (see Table~1) from the volume-limited
representative sample of local LIRGs of Alonso-Herrero et
al. (2006). This sample was 
drawn from the {\it IRAS} Revised Bright Galaxy Sample (RBGS, Sanders et
al. 2003). The Alonso-Herrero et al. (2006) sample is limited in distance  
(velocities of $2750-5200\,{\rm km\,s}^{-1}$ or distances of 
$d\sim 35-75\,$Mpc for
the assumed cosmology) so that the  
Pa$\alpha$ emission line at rest-frame wavelength $1.875\,\mu$m could be
observed with the NICMOS F190N narrow band 
filter (see Sect.~2.3 and Alonso-Herrero et al. 2000a, 2006 for full
details). Arp~299, the
most luminous system in this LIRG sample, was observed with the IFS
INTEGRAL instrument and analyzed in detail by 
Garc\'{\i}a-Mar\'{\i}n et al. (2006). For the sake of 
completeness we will include 
Arp~299 ($ \log (L_{\rm IR}/{\rm L}_\odot) =11.88$) 
in the discussions presented in Sects.~4, 5, and 6. 
Our sample of LIRGs covers a range in infrared
luminosities of  
$ \log (L_{\rm IR}/{\rm L}_\odot) =11.05-11.88$. 
The average infrared luminosity of the full sample
(northern and southern hemispheres) is $L_{\rm IR}=2.1 \times
10^{11}\,{\rm L}_\odot$ or $ \log (L_{\rm IR}/{\rm L}_\odot) =11.32$.
Except for Arp~299, which is a
strongly interacting system, the rest of
the northern hemisphere LIRGs in this sample are apparently isolated galaxies,  
weakly interacting systems (e.g., NGC~7469, NGC~7771) and galaxies
with small companions but no clear 
morphological signs of interaction (e.g., NGC~6701, see M\'arquez et
al. 1996).

\subsection{PMAS Optical Integral Field Spectroscopy Observations}

We obtained  optical IFS of 11 LIRGs using PMAS 
on the 3.5\,m telescope at the German-Spanish
Observatory of Calar Alto
(Spain) during three observing runs: November 2005,
May 2006, and December 2006. The PMAS observations were taken with  
the Lens Array Mode configuration which is
made of  a $16 \times 16$ array of microlenses coupled with fibers 
called hereafter spaxels. We used the 1\arcsec \, magnification which
provides a field
of view (FoV) of $16\arcsec \times 16\arcsec$.  We used the V300 grating with 
a dispersion of $1.67\,$\AA\, pixel$^{-1}$ and an approximate spectral range
of 3400\,\AA. The wavelength range covered by the observations was
approximately $3800-7200\,$\AA.
The approximate spectral resolution of the spectra in
the $2\times 2$ binned mode is 
6.8\,\AA  \ full width half maximum (FWHM, see also Sect.~3.1).

\begin{figure}
\setcounter{figure}{1}

\resizebox{\hsize}{!}{\includegraphics[width=3cm]{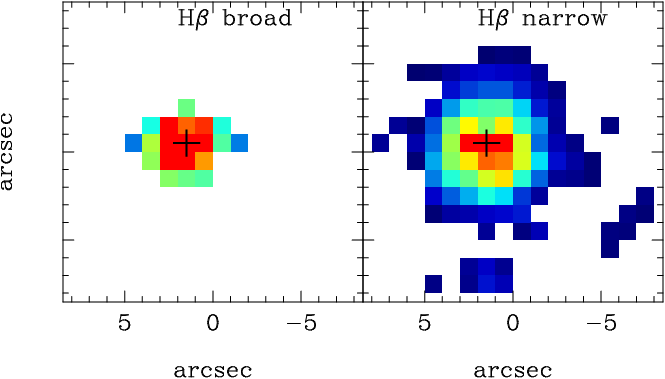}}

\vspace{0.5cm}
\resizebox{\hsize}{!}{\includegraphics[width=3cm]{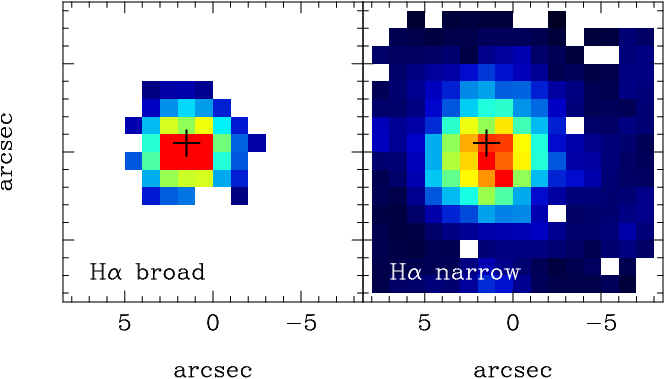}}

\caption{PMAS maps of the H$\beta$ and H$\alpha$ emission lines of 
NGC~7469 fitted using two components, broad (left panels) and narrow
(right panels). Orientation is north up, east to the left.
The images are shown on a square root scale.}
\end{figure} 

\subsubsection{Observing procedure}

The total integration time for each galaxy was split into three or four
individual galaxy exposures (see Table~1), 
each of which having between 400 and 1000\,s.   
Given the relatively small FoV of PMAS in the Lens Array configuration and the
large extent of the galaxies, we obtained a separate  sky integration for each
galaxy, interleaved with the galaxy observations and with a comparable
single exposure time of  
900 or 1000\,s. For all the galaxies we obtained one pointing, except for 
NGC~7771 for which we took two pointings to cover the approximate
central $28\arcsec \times 16\arcsec$ region.

Table~1 gives for each target the observing campaign, integration times for the
individual galaxy exposures and number of exposures, 
integration time, 
seeing conditions for each data set, as well as the airmasses
of the observations.  The typical seeing conditions of the observations 
varied between 0.8\arcsec \, and 1.7\arcsec, 
depending on the observing campaign. 

During each night we obtained
calibration arcs and lamps, and flat-fields.
Since PMAS is located at the Cassegrain focus, it is affected
by the changing 
flexures of the telescope when tracking the targets. 
Therefore, we obtained individual arc and
internal lamps exposures for each different pointing, when applicable before
and after culmination of the target. The lamps used were a ThAr lamp 
for the Nov 2005 and May 2006 campaigns, and a HgNe lamp for the Dec
2006 campaign.
 Additionally throughout the nights we observed spectrophotometric
standard stars to correct for the instrument response, and to flux calibrate
the data. We note, however  that conditions were
non-photometric for several nights in all the campaigns (see Table~1).

\subsubsection{Data reduction}

The PMAS data were reduced using a set of customized scripts under the
IRAF\footnote{IRAF software is distributed by the National Optical Astronomy
Observatory (NOAO), which is operated by the Association of Universities for
Research in Astronomy (AURA), Inc., in cooperation with the National Science
Foundation.} environment.
The first step of the data reduction was to determine and subtract the bias
level. Next, using reference internal continuum lamp exposures, we
identified and traced the location of each of the 
256 spectra along the dispersion
direction of the CCD. Once we extracted the individual spectra, the third
reduction step was the wavelength calibration, which we carried out using a
model obtained from arc lamp exposures. 
We checked the wavelength calibration against known sky emission lines
and measured a median standard deviation of 1.8, 2.0, and 1.8\AA\, for the
Nov 2005, May 2006,  and Dec 2006 observing runs, respectively. The fourth step
was to correct for the sensitivity variations by creating a response image
using internal calibration lamp images and a sky flat exposure. After that, a
relative flux calibration was performed using standard star observations. The
next step was to combine the different galaxy exposures (a  minimum of three)
for each individual pointing to improve the signal-to-noise ratio and 
to remove  cosmic rays. The sky subtraction was done on
a spaxel-by-spaxel basis using for each galaxy its own sky
observation. 
The final step was to build the data cubes and rotate them to the north
up, east to the left orientation.

\begin{table*}
\begin{center}
\caption{Nuclear and Integrated observed (not corrected for
  extinction) line ratios.}
\begin{tabular}{lcccccc}
\hline
Galaxy        & Type        & Size & [OIII]/H$\beta$ &[OI]/H$\alpha$ &[NII]/H$\alpha$
&[SII]/H$\alpha$ \\
(1) & (2) & (3) & (4) & (5) & (6) & (7) \\
\hline
\hline
NGC~23	      & Nuclear     & 0.28   &1.52      &0.091   &0.87
&0.48       \\
	      & Integrated  & 4.5    &0.58      &0.044   &0.57
              &0.33       \\
\\
MCG~+12-02-001 & Nuclear     & 0.31   &0.67      &0.029   &0.44
&0.30      \\ 
	      & Integrated  & 5.8    &0.61      &0.030   &0.42
              &0.35      \\ 
\\
UGC~1845	      & Nuclear     & 0.30   &2.07      &0.090   &1.09
&0.33      \\ 
	      & Integrated  & 3.2    &1.33      &0.065   &0.72
              &0.28      \\ 
\\
NGC~2388	      & Nuclear     & 0.28   &0.32      &0.029   &0.63
&0.22      \\
	      & Integrated  & 5.1    &0.60      &0.066  &0.56
              &0.30      \\
\\
MCG~+02-20-003 & Nuclear     & 0.33   &0.67      &0.054   &0.63
&0.37      \\ 
	      & Integrated  & 6.5    &0.65      &0.046   &0.45
              &0.43      \\ 
\\
IC~860	      & Nuclear     & 0.29   & $\cdots$  & $\cdots$     &7.81
&3.85      \\
\\
NGC~5936	      & Nuclear     & 0.30   &0.32	&0.031   &0.63
&0.22      \\
	      & Integrated  & 4.7    &0.24      &0.034   &0.48
              &0.25      \\  
\\
NGC~6701	      & Nuclear     & 0.27   &0.68      &0.065   &0.72
&0.35      \\  
	      & Integrated  & 5.2    &0.62      &0.054   &0.67
              &0.35      \\ 
\\
NGC~7469$^{*}$	      & Nuclear     & 0.32   &9.96      &0.13    &4.10
&0.39      \\
                      & Integrated & 6.1   & 3.53   & 0.048 & 0.55 & 0.30
\\
\\
NGC~7591	      & Nuclear     & 0.32   &0.98      &0.11    &0.96
&0.41      \\      
	      & Integrated  & 5.1 & 0.85 & $\cdots$ & 0.85 & 0.59 \\
\\
NGC~7771	      & Nuclear     & 0.28   &0.22      &0.032   &0.40
&0.21      \\
	      & Integrated  & 8.0    &0.42      &$\cdots$      &0.55
              &0.43  \\ 

\hline
\end{tabular} 
\end{center}
Notes.--- Column~(2): PMAS type of 1D spectra.
Column~(3): Linear physical size in kpc
covered by the PMAS extraction apertures for the 
nuclear and  integrated spectra using the distances
given by Sanders et al. (2003) for the {\it IRAS} RBGS. Columns~(4), (5),
(6), and (7):
Observed (not corrected for extinction and/or stellar absorption) line
ratios. $^{*}$For NGC~7469 the line ratios correspond to the narrow components
of the hydrogen recombination lines.

\end{table*}

\subsection{{\it HST}/NICMOS Observations}
The {\it HST}/NICMOS observations were obtained with the NICMOS NIC2
camera, which has a pixel size of 0.076\arcsec \, and a FoV of 
$\sim 19.5\arcsec \times 19.5\arcsec$. 
Details on the observations and data reduction procedures can
be found in  Alonso-Herrero et al. (2006). For this paper we make use
of the $1.6\,\mu$m continuum observations and  
the continuum-subtracted Pa$\alpha$ images for comparison with the PMAS data. 
The only additional step needed for the NICMOS images was to rotate
and trim them 
to match the orientation and FoV, respectively, of the PMAS
images. The angular resolution of the NICMOS data is approximately 
$0.15\arcsec$ and $0.18\arcsec$ (FWHM) for the $1.6\,\mu$m continuum
and Pa$\alpha$ images, respectively.

\section{IFS Data Analysis}
 
\subsection{Spectral maps}

We constructed spectral maps of the brightest emission lines by
fitting the lines to Gaussian functions and the adjacent 
continuum to straight
lines, on a spaxel-by-spaxel basis. To do so we developed our own
routines which make use of the IDL-based MPFITEXPR 
algorithm\footnote{http://www.purl.com/net/mpfit} 
developed by Markwardt (2008)
to measure in an automated fashion the  central wavelength, the width of 
the Gaussian ($\sigma = {\rm FWHM}/2.35$), and
integrated flux for each emission line. Using this algorithm we were
also able to fix  the relative wavelengths and  line 
ratios according to atomic parameters when fitting multiple emission
lines ([N\,{\sc ii}]$\lambda$6548, H$\alpha$, [N\,{\sc
  ii}]$\lambda$6584, 
and [S\,{\sc ii}]$\lambda$6717,6731). 
Additionally for each of the two sets of lines we imposed
that the lines had the same width. 
We did not attempt to correct the maps of the H$\beta$ emission line
for the presence of H$\beta$ in absorption, which is observed in
most of the spectra of the sample of LIRGs (see also Sect.~3.2).

\begin{figure*}
\setcounter{figure}{2}


\includegraphics[width=13cm]{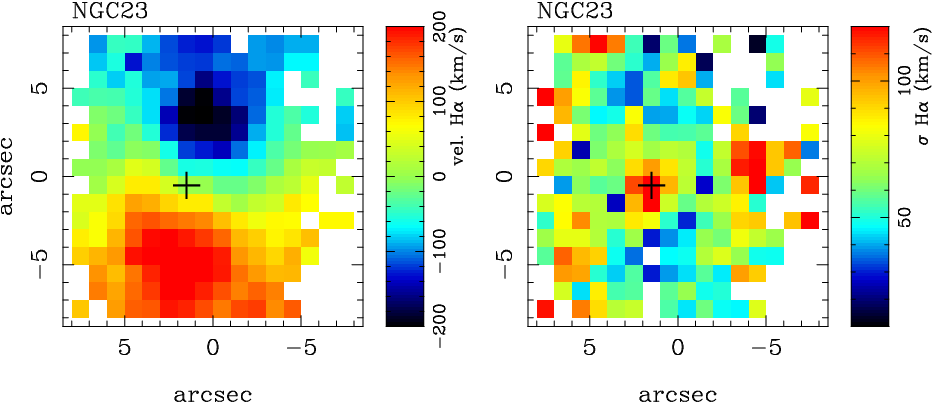}

\includegraphics[width=13cm]{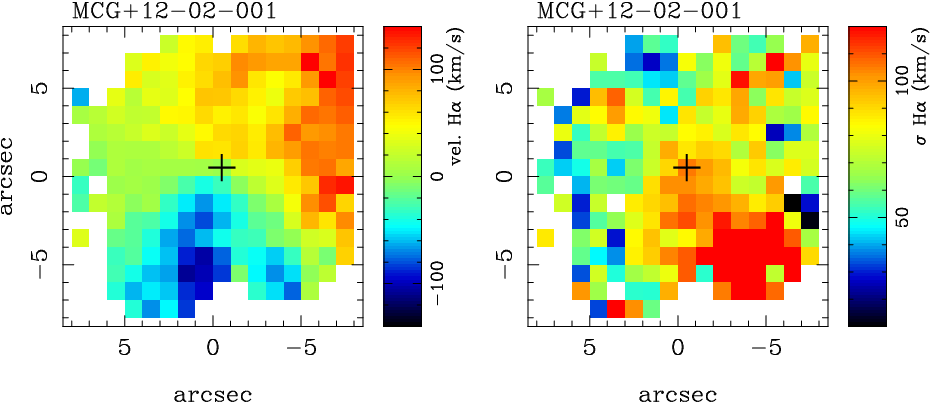}

\includegraphics[width=13cm]{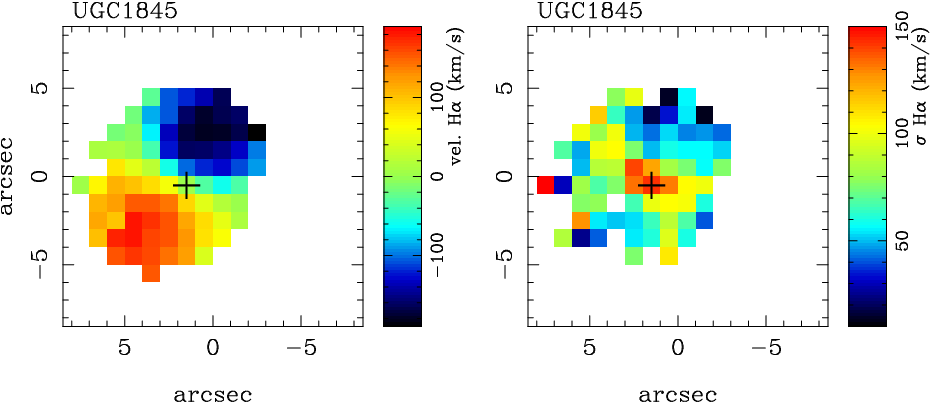}
\vspace{0.1cm}
\caption{(a) Maps of the observed H$\alpha$ velocity field (left panels) 
and the velocity 
dispersion (right panels). The latter has been corrected for the instrumental 
resolution. The zero points for the H$\alpha$ velocity fields
are set at the location of the peak of the 6200\,\AA \, continuum
emission as marked with the crosses
for each galaxy. Orientation is north up, east to the left.} 

\end{figure*}

\begin{figure*}
\setcounter{figure}{2}

\includegraphics[width=13cm]{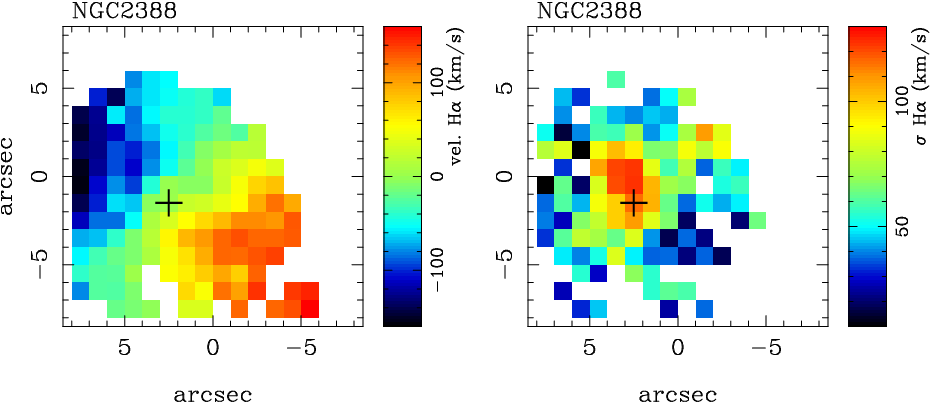}

\includegraphics[width=13cm]{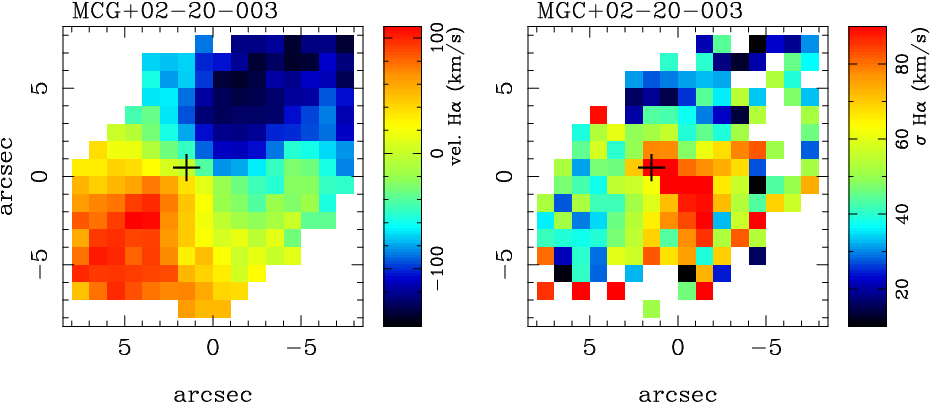}

\includegraphics[width=13cm]{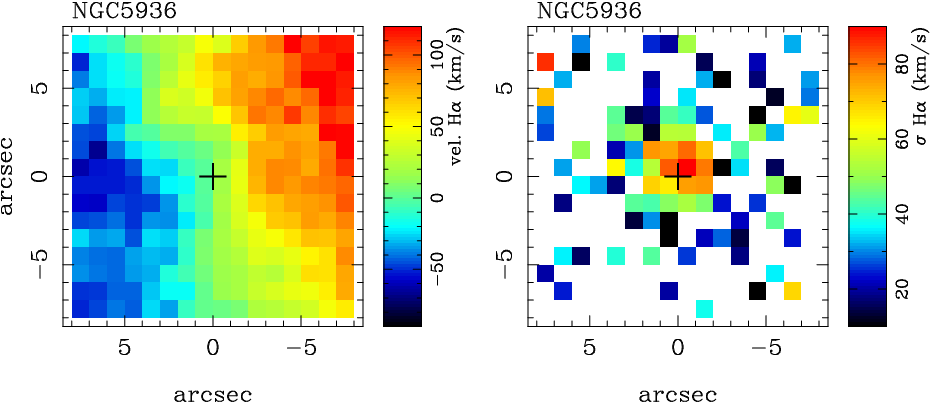}
\vspace{0.1cm}
\caption{Continued. } 
\end{figure*}

\begin{figure*}

\setcounter{figure}{2}

\includegraphics[width=13cm]{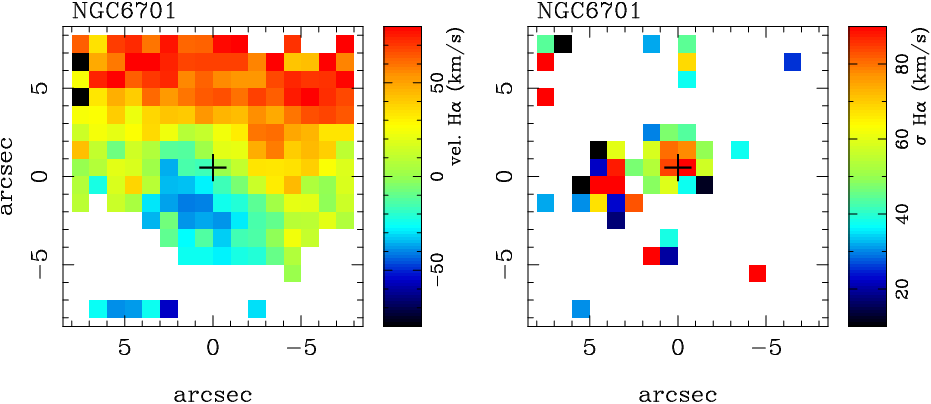}

\includegraphics[width=13cm]{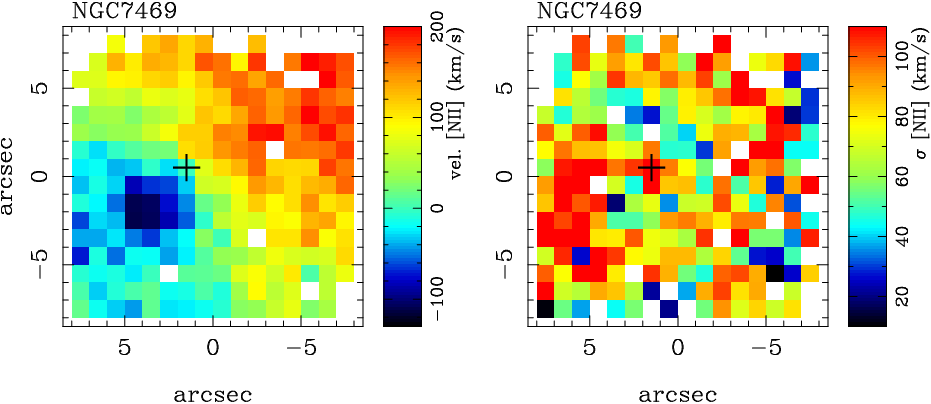}

\includegraphics[width=13cm]{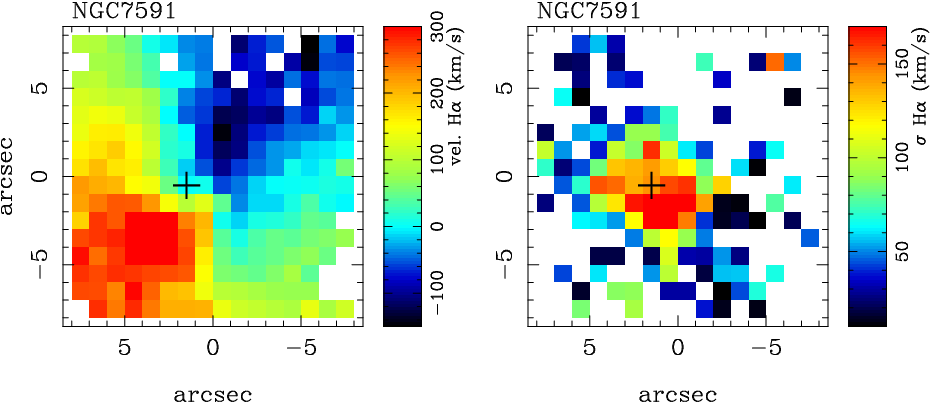}

\vspace{0.1cm}
\caption{(b) As Fig.~3a. The velocity field and the map of the velocity
dispersion for NGC~7469 are for the [N\,{\sc ii}]$\lambda$6584
emission line instead of H$\alpha$.} 

\end{figure*}

The spectral maps of the emission line fluxes, as well as the maps of
the central wavelength (or velocity field), and $\sigma$ (or velocity
dispersion)  were only constructed for
those spaxels whose integrated line flux was $3\,\delta$ above the
local continuum, where $\delta$ was the standard deviation of the 
fitted local continuum.  
We found that the emission lines of  all galaxies were adequately
fitted with one component (Fig.~1), except for NGC~7469, a
well-known Seyfert 1 galaxy, for which we fitted the hydrogen 
recombination lines using a broad and a narrow component. The maps of
both components are shown in Fig.~2.

The uncertainties of the measured
velocities and velocity dispersions depend on 
the errors in measuring the centroid and width of the emission line, which in
turn depend on 
the signal-to-noise of
the spaxel, and the systematic errors (e.g., the wavelength
calibration,
see Sect.~2.2). The typical errors in measuring the widths of the
Gaussians when fitting the H$\alpha$+[N\,{\sc ii}] lines were 10\%.
The maps of the observed H$\alpha$ velocity dispersion 
($\sigma_{\rm obs}$) were corrected for the instrumental 
resolution ($\sigma_{\rm ins}$) 
using  $\sigma^2 = \sigma_{\rm
  obs}^2-\sigma_{\rm ins}^2$. For each observing campaign
we measured the instrumental resolution  on a spaxel-by-spaxel
basis using the observations of the arc emission lines at a wavelength 
close to that of H$\alpha$. The median instrumental resolution across
the detector near H$\alpha$ is $\sigma_{\rm
  ins}=2.9\pm 0.2\,$\AA \, for all three observing runs. 
The H$\alpha$ velocity fields and the maps of the
H$\alpha$ velocity dispersion are shown in the left and right panels,
respectively, of Fig.~3. 

Since the  emission lines observed  in IC~860 are faint, we decided to fit
the H$\alpha$+[N\,{\sc ii}] and the [S\,{\sc ii}] lines (H$\beta$, 
[O\,{\sc iii}]$\lambda$5007 and [O\,{\sc i}]$\lambda$6300 were not detected) 
manually using {\it splot} within {\sc iraf} without any
constraints. The spectral maps for this galaxy are shown in
Fig.~1f. As can be seen from this figure, the
H$\alpha$ and [S\,{\sc ii}] lines are only
detected in the nuclear regions, whereas the [N\,{\sc
  ii}]$\lambda$6584 emission line is more extended.

In addition to fitting the brightest emission lines for each galaxy,
we constructed a continuum image centered at $\lambda \simeq 6200\,$\AA \ by
summing up the continuum spectra over a rectangular band width of 
approximately 17\,\AA. Fig.~1 shows the 
continuum maps for our sample of LIRGs. The peak of the
optical continuum emission is marked with a cross for each galaxy 
  on the PMAS maps in this
figure as well as in Figs.~2 and 3.

As explained in Sect.~2.1.1, for 
NGC~7771 we took two separate pointings to cover the approximate
central $28\arcsec \times 16\arcsec$ region of the galaxy. 
The systematic error of telescope offsets with PMAS is given by the
astrometric accuracy of the PMAS Acquisition and Guiding (A\&G) system, which has been determined
to be  $0.1966 \pm 0.0004\arcsec$/pixel (Roth et al. 2005). As the A\&G
system is an integral part of the instrument, systematic offset errors
are negligible. The statistical accuracy of the auto-guider is a fraction
of a pixel, and the canonical value is $0.1\arcsec$ r.m.s. (Roth, 2009, private
communication). Thus, we used the offsets commanded using the PMAS
acquisition images to construct mosaics of the emission line and
continuum maps, as well 
as the H$\alpha$ velocity fields and maps of the velocity dispersion.

Fig.~1 shows for the 11 LIRGs in our sample the PMAS maps of the
brightest optical emission lines, the {\it HST}/NICMOS continuum
subtracted Pa$\alpha$ 
images, together with images of the  
stellar emission at $\sim 6200\,$\AA \, (PMAS) and at $1.6\,\mu$m
(NICMOS). Since the PMAS images do not have astrometry, the PMAS and
NICMOS images were registered by using the peaks and shapes of the
continua for reference. For the typical distances of  
our galaxies the FoV of the PMAS observations cover the central
$4.3-5.3\,$kpc, except for NGC~7771 for which the PMAS mosaics cover
approximately the central $7.8\,{\rm kpc} \times 4.5\,{\rm kpc}$.
The maps are shown on a square root scale to maximize the contrast
between diffuse and bright regions.

The observed H$\alpha$ velocity fields for all the LIRGs in our sample
except for IC~860 are shown in Fig.~3. For 
NGC~7469 we show the  [N\,{\sc ii}]$\lambda$6584 velocity field
instead, as that of H$\alpha$ is affected by the uncertainties associated
  with fitting the broad and narrow components.
The zero points of the 
velocity fields are set at the peak of the 6200\,\AA \, continuum
emission. The only exception is NGC~7771 where the zero point
is placed at the position that makes the velocity field gradient
symmetric, and it is approximately coincident with the peak of the
near-infrared continuum.
  Table~3 gives the measured H$\alpha$ $cz$ for the nuclei of
our sample of LIRGs.

\subsection{Extraction of 1D spectra}

For each galaxy we extracted two 1D spectra: the nuclear spectrum and
the integrated spectrum.  The nuclear spectra correspond to the 
spaxel at the peak of the 6200\,\AA \, continuum emission. The
physical size covered by the nuclear spectrum is given for each galaxy
in Table~2, and it is typically the approximate central 300\,pc. 
For reference the nuclear spectra of
Kim et al. (1995) and Veilleux et al. (1995) 
were extracted with a linear physical size of 
2\,kpc (see discussion in Sect.~5). 

The integrated spectrum of each galaxy was extracted by defining
$\sim 6200\,$\AA \, continuum 
isophotes and then summing up all the spaxels contained within the
chosen external continuum isophote. The external isophotes (plotted
for all galaxies in Fig.~1) were selected  
to cover the PMAS FoV  
as much as possible, without compromising the quality
of the extracted 
spectra. In Table~2 we  give for each galaxy the
approximate physical size along the major axis of the galaxy of the
outer continuum isophote used for  the
integrated spectra. Note that the term integrated is used in the sense
of integrated  
spectra over the PMAS FoV, and the integrated spectra do not encompass the whole galaxy 
(see e.g. figure.set.8 of Moustakas \& Kennicutt 2006 for a comparison 
with one of the galaxies in our sample, NGC~23). Typically our integrated 
spectra cover the central 
3 to 8\,kpc along the major axis of the galaxy, depending on the galaxy, but for most galaxies 
they include the central $\sim 5\,$kpc (see Table~2).  
The integrated spectra of those galaxies
observed under non-photometric conditions (Table~1) may be 
affected by imperfect sky
subtraction, but this does not affect the measurements of the brightest
emission lines. 
The full nuclear spectra are shown in Fig.~4 for each LIRG
in the sample. In the same figure we present in the insets
 the blue part of the integrated spectra to
emphasize the differences between the underlying absorption features and the
H$\beta$ and [O\,{\sc iii}]$\lambda$5007 emission lines in the nuclear
and integrated spectra.  

The fluxes of the brightest emission lines of the nuclear and
integrated spectra were measured manually using {\it splot} within
{\sc iraf} and were compared with the automated flux measurements used for
constructing the spectral maps (see Sect.~3.1), except for NGC~7469. 
In the case of the manual measurements we did not impose any
constraints on the line widths and ratios 
when measuring the [N\,{\sc ii}]$\lambda$6548, H$\alpha$ and
[N\,{\sc ii}]$\lambda$6584, and the [S\,{\sc ii}]$\lambda \lambda
$6717,6731 lines. For the nuclear and integrated 
spectra of NGC~7469 we used the method and restrictions
described in Sect.~3.1.
We find  a good agreement between the manual and the automated
measurements of the nuclear values. The largest differences (up to $\sim
30\%$)  are for the [O\,{\sc i}]$\lambda$6300/H$\alpha$ line ratio,
whereas for the other line ratios the differences are always of less
than $15\%$. As done for the spectral maps,
we did not attempt to correct for the presence of H$\beta$  
in absorption. The observed (not corrected for extinction) line ratios
for the nuclear and integrated  spectra are given in Table~2.

\section{Morphology of the stellar and gas emissions}

The optical and near-infrared continuum images (Fig.~1)
reveal the presence of bright nuclei, and a large 
number of 
star clusters in the nuclear regions as well as along the large scale
spiral arms. The star clusters are only unveiled by the higher angular 
resolution of the NICMOS images, which is
typically a few tens of parsecs for our sample of
LIRGs. There is a good overall correspondence, on scales of a few hundred
parsecs (the PMAS spatial resolution),  between the optical and the
near-infrared stellar continua as mapped out by the PMAS 6200\,\AA \ and the
NICMOS $1.6\,\mu$m emissions, respectively.  This indicates that 
for the majority of these LIRGs the effects of extinction on the
continuum morphologies are not
overly severe, except for the nuclei of the galaxies 
(Alonso-Herrero et al. 2006). 
This is clearly seen in 
NGC~7771 where the peaks of the optical and near-infrared continua
appear displaced by a few arcseconds (see Figs.~1k and 3). This is probably 
due to the highly inclined nature of this galaxy, as well as the
diversity of stellar populations and patchy extinction present in the
ring of star formation (Davies, 
Alonso-Herrero \& Ward 1997; Smith et al. 1999; Reunanen et al. 2000).
Another example with large differences between the 
optical and near-infrared continuum emission is Arp~299, and we 
refer the reader to Alonso-Herrero et al. (2000a) and
Garc\'{\i}a-Mar\'{\i}n et al. (2006) for a full discussion.

The PMAS H$\alpha$ and the NICMOS
Pa$\alpha$ emissions are well correlated, and  both trace  the
nuclear emission as well as the emission from bright, high 
surface-brightness H\,{\sc ii } regions (Fig.~1). The high angular resolution 
of the NICMOS
Pa$\alpha$ images resolves with exquisite detail the sites of the youngest
star forming regions in the central regions of LIRGs. These high
surface-brightness H\,{\sc ii} regions can be either 
located in the central $1-2\,$kpc
or spread out throughout the disk of the galaxies 
(see Alonso-Herrero et al. 2006 for more details).  

Since the NICMOS Pa$\alpha$ and nearby continuum 
images were taken with narrow-band filters,
and a relatively small pixel size, they are not very sensitive to the diffuse
low surface-brightness  emission. This may also be due to the fact that 
in LIRGs the diffuse emission suffers much less extinction than the bright 
H\,{\sc ii} regions (see Rieke et al. 2009 and references therein). 
The PMAS H$\alpha$ images show emission from
 the H\,{\sc ii} regions, at lower angular resolution than the NICMOS
 images.  Additionally, for most LIRGs the 
PMAS H$\alpha$ images are also sensitive to more diffuse low 
surface-brightness 
emission. This  extended  emission can be seen almost over 
the entire PMAS FoV (e.g., NGC~5936
 Fig.~1g and NGC~6701 Fig.~1h), and beyond, as shown by other works 
for a few  galaxies in common with our sample (M\'arquez, Masegosa, \& Moles
 1999; Dopita et al. 2002; Hattori et al. 2004).  
It is only the  two galaxies with the most compact nuclear Pa$\alpha$ 
emission, IC~860 (Fig.~1f) and to a lesser degree UGC~1845 (Fig.~1c), 
that also show relatively compact H$\alpha$ emission.

The fact that H$\alpha$ and Pa$\alpha$ morphologies are 
in general similar suggests that over the PMAS FoV and with the PMAS 
angular resolution the extinction effects on H$\alpha$ are not severe,
except
in the very nuclear regions. Indeed, Alonso-Herrero et
al. (2006) for the galaxies in common with this work measured 
average extinctions to the gas of $A_V\sim 2-4\,$mag over the Pa$\alpha$
emitting regions (a few kpc). Similar values of the extinction were 
found by Veilleux et al. (1995) from the Balmer decrement. 
The PMAS H$\beta$ maps are similar to those of H$\alpha$ although they
are more affected by extinction and/or  the presence of 
H$\beta$ in absorption, 
especially in the
nuclear regions (e.g., UGC~1845 Fig.~1c, MCG+02-20-003 Fig.~1e, NGC~7591 Fig.~1j).

\begin{figure*}

\setcounter{figure}{2}

\includegraphics[width=5.cm,angle=-90]{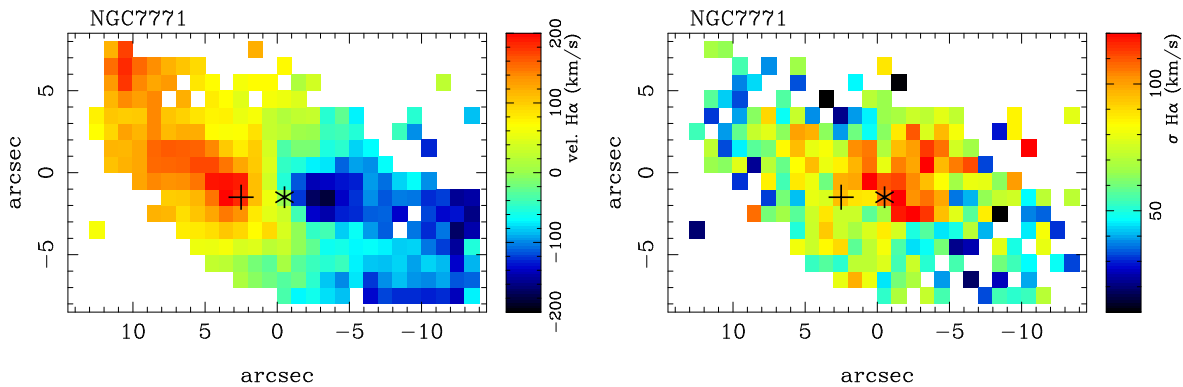}

\caption{(c) As Fig.~3a. In the case of NGC~7771 the 
asterisk marks the position that
  makes the velocity field gradient symmetric, and it is approximately
  coincident with the peak of the near-infrared continuum and the maximum of
the velocity dispersion.} 
\end{figure*}

In general the overall morphology of the brightest forbidden lines ([N\,{\sc
  ii}]$\lambda 6584$,  [S\,{\sc ii}]$\lambda\lambda$6717,6731) shows a
reasonable correlation with that of H$\alpha$. On  smaller scales,
however, 
some differences are already apparent in Fig.~1.  For instance, in NGC~23
(Fig.~1a) all the optical 
emission lines except for H$\alpha$ and H$\beta$ peak in the 
nucleus, whereas the brightest regions of hydrogen recombination line 
emission (H$\alpha$, Pa$\alpha$), which trace the sites of on-going 
star formation, are in the circumnuclear ring of star
formation. There are also small scale differences between the
H$\alpha$ and [N\,{\sc
  ii}]$\lambda 6584$ emissions of the nuclei and the H\,{\sc ii} regions
For example, 
some of the extra-nuclear H\,{\sc ii} regions of NGC~2388 (Fig.~1d)
and MCG~+02-20-003 (Fig.~1e) appear to have lower 
[N\,{\sc  ii}]$\lambda 6584$/H$\alpha$ ratios than their nuclei. This
is a well known behavior of galaxies (Veilleux \& Osterbrock 1987;
Kennicutt, Keel, \& 
Blaha 1989; Sarzi et al. 2007). 
The small scale differences of the different emission lines
in LIRGs are 
more apparent from the spatially resolved properties of the optical
line ratios of this sample of LIRGs  and will be discussed in detail 
in a forthcoming paper (Alonso-Herrero et
al. 2009, in preparation).

Although the limited angular resolution of the PMAS continuum and gas
maps does not allow us to resolve all the morphological details seen
in the NICMOS images, the
general behavior is the same. For the majority of the LIRGs in
our sample the peaks of the stellar and the H$\alpha$ gas emissions are
coincident and located in the nuclear region. The few cases of displacements
between the stellar and the H$\alpha$ peaks are found in those 
LIRGs with circumnuclear rings of star formation without Seyfert activity. In 
NGC~23 (Fig.~1a) and NGC~7771 (Fig.~1k) the stellar emission
peaks in the nuclei, whereas the brightest H$\alpha$ and Pa$\alpha$ emission are
in luminous H\,{\sc ii} regions in the rings. These displacements 
are likely to be due to differences in the stellar populations.
That is, the youngest
regions and thus brightest H$\alpha$ emitting regions are in the rings,  
whereas the nuclei contain older stellar populations. The latter is
clearly demonstrated by the presence of strong absorption features 
(Balmer line series, H+K CaII lines) in the nuclear spectra 
of these two galaxies (Fig.~5). In the case of NGC~7469, 
which also shows a bright circumnuclear ring of star formation (see
Genzel et al. 1995; D\'{\i}az-Santos et al. 2007, and references
therein), the peaks of the
stellar and gas emission are coincident with the nuclear AGN.  

In contrast to the majority of the LIRGs in this volume-limited sample, 
in the interacting LIRG Arp~299 
 the peaks of the observed warm ionized gas are
displaced from the stellar peaks typically by $\sim 1.4\,$kpc
(see  Garc\'{\i}a-Mar\'{\i}n et al. 2006). In local interacting ULIRGs 
the displacements between the continuum and gas emissions are common
and even larger, typically $2-4\,$kpc and in some exceptional cases as large
as $\sim 8\,$kpc 
(Colina et al. 2005; Garc\'{\i}a-Mar\'{\i}n et al. 2009). The
differences in the stellar and gas distributions in ULIRGs 
are understood in terms of 
the more extreme effects produced by the interaction processes on the
spatial distribution of the ionizing sources and the presence of large
amounts of dust in the nuclear regions.

\section{Nuclear versus integrated spectral classification}
We used the standard optical diagnostic diagrams (BPT diagrams,
Baldwin, Phillips, \& Terlevich 1981) to 
classify galaxies into the H\,{\sc ii}-like, LINER, and Seyfert
spectral types. For the spectral classification of the nuclear
  and integrated activity 
we used two different sets of boundaries in the optical line ratio
diagrams. The first are the classical semi-empirical boundaries of Veilleux \&
Osterbrock (1987, V\&O87 hereafter), which have the added advantage of making
the comparison with  previous results on these LIRGs easier. 
The V\&O87 boundaries are shown in Fig.~5. The second set includes the
latest empirical and theoretical boundaries derived by  
Kauffmann et al. (2003) and Kewley et al. (2001a, b; 2006), and are
shown in Fig.~6. Kewley et al. (2001a, b) modeled AGN
and starburst line ratios to 
provide theoretical boundaries in these diagnostic diagrams.  
They defined the so-called maximum starburst lines 
above which the line ratios cannot be explained by pure star formation. 
Kauffmann et al. (2003) and Kewley et al. (2006)
derived empirical boundaries in these diagrams to separate H\,{\sc ii},
LINERs and Seyfert galaxies using large samples of galaxies drawn from  the
Sloan Sky Digital Survey (SSDS). We will refer to this second set of
boundaries as theoretical/SSDS.

\begin{figure*}

\setcounter{figure}{3}

\includegraphics[width=6.5cm,angle=-90]{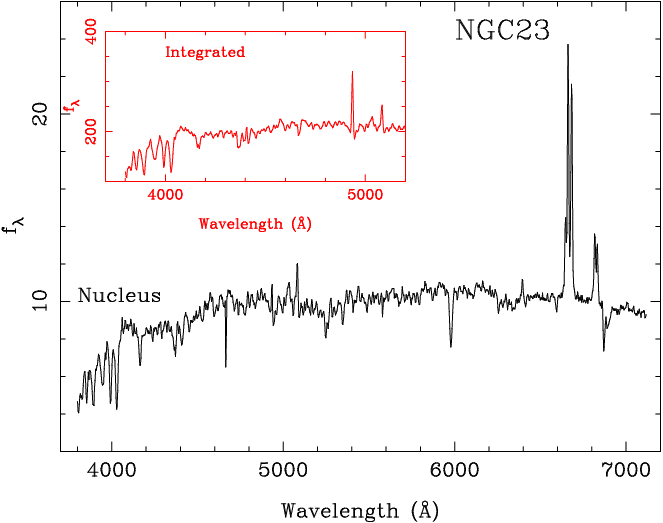}
\includegraphics[width=6.5cm,angle=-90]{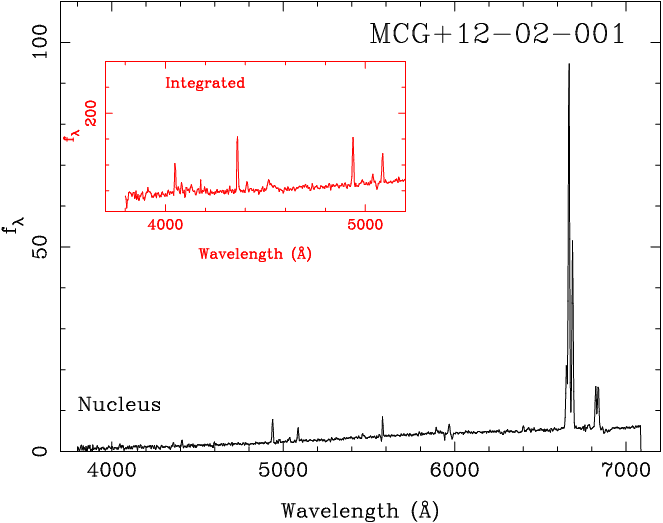}

\vspace{0.5cm}
\includegraphics[width=6.5cm,angle=-90]{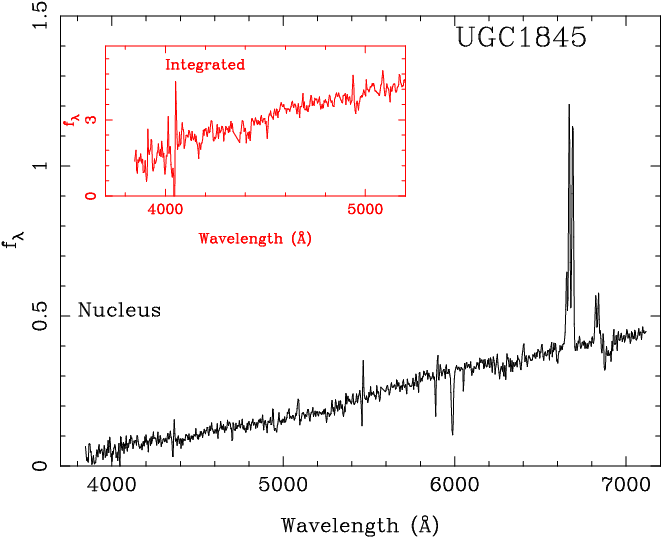}
\includegraphics[width=6.5cm,angle=-90]{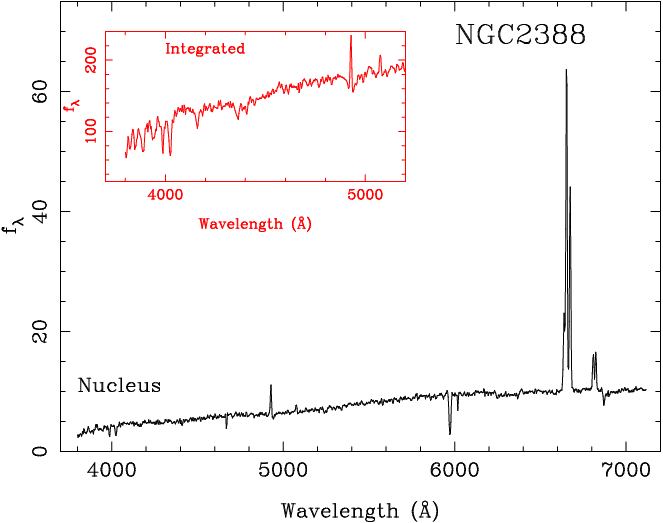}

\vspace{0.5cm}
\includegraphics[width=6.5cm,angle=-90]{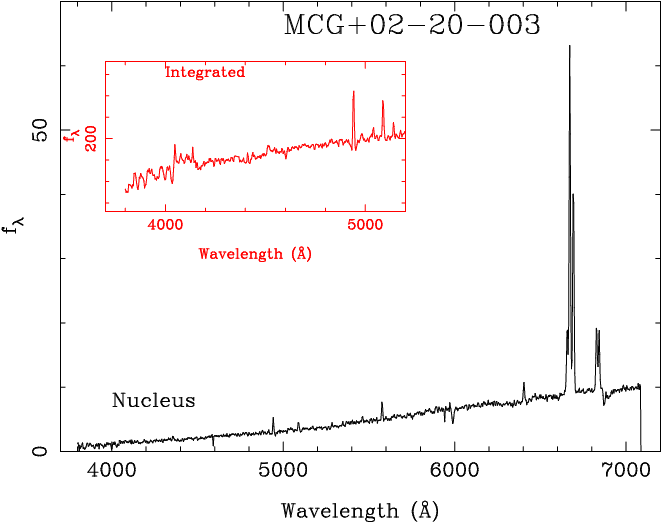}
\includegraphics[width=6.5cm,angle=-90]{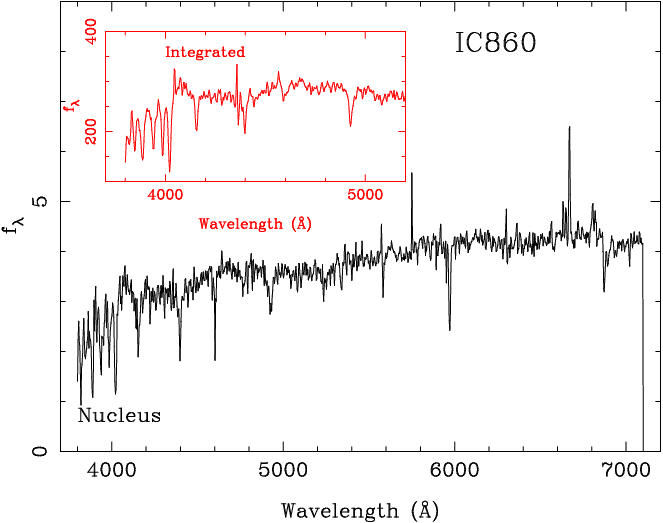}

\caption{PMAS spectra (plotted in arbitrary units) 
of the nuclei of the galaxies in the
  sample. Additionally, for each LIRG 
the inset shows the blue part of the integrated spectrum 
so that the absorption features as well as the H$\beta$ and 
the [O\,{\sc iii}]$\lambda$5007 emission lines can be clearly seen. 
In the case of NGC~7469 we also show in the lower right panel 
the full spectrum of an H\,{\sc ii} region located in the
circunmnuclear ring of star formation.} 
\end{figure*} 

\begin{figure*}

\setcounter{figure}{3}

\vspace{0.5cm}
\includegraphics[width=6.5cm,angle=-90]{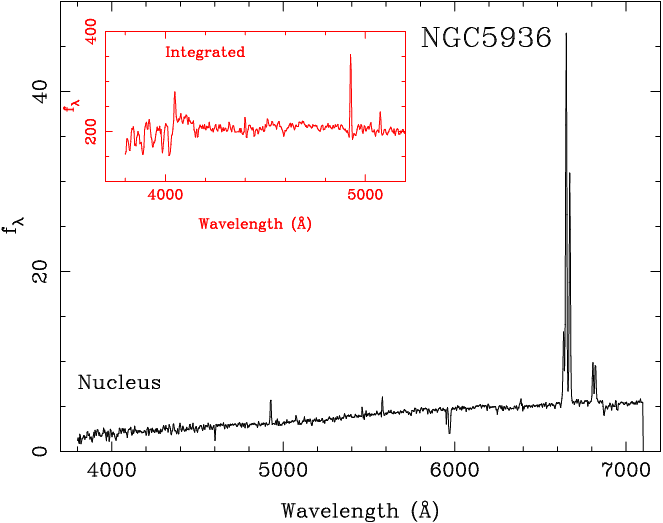}
\includegraphics[width=6.5cm,angle=-90]{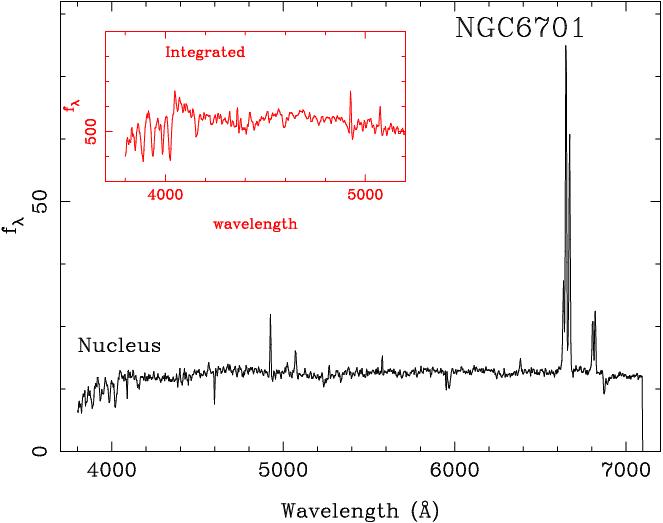}

\vspace{0.5cm}
\includegraphics[width=6.5cm,angle=-90]{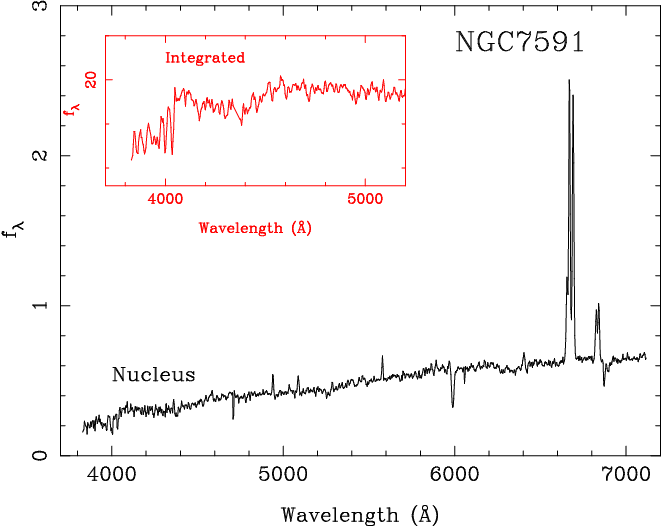}
\includegraphics[width=6.5cm,angle=-90]{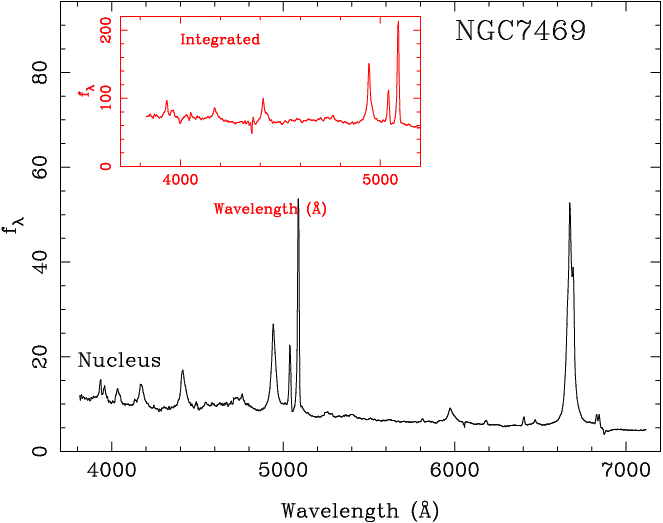}

\vspace{0.5cm}
\includegraphics[width=6.5cm,angle=-90]{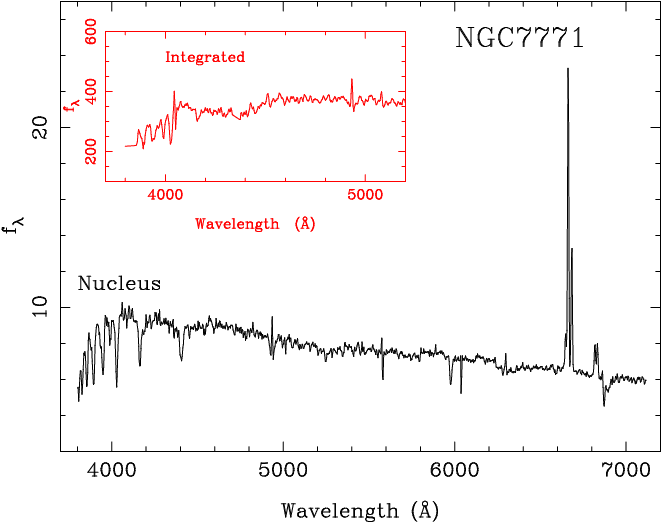}
\includegraphics[width=6.5cm,angle=-90]{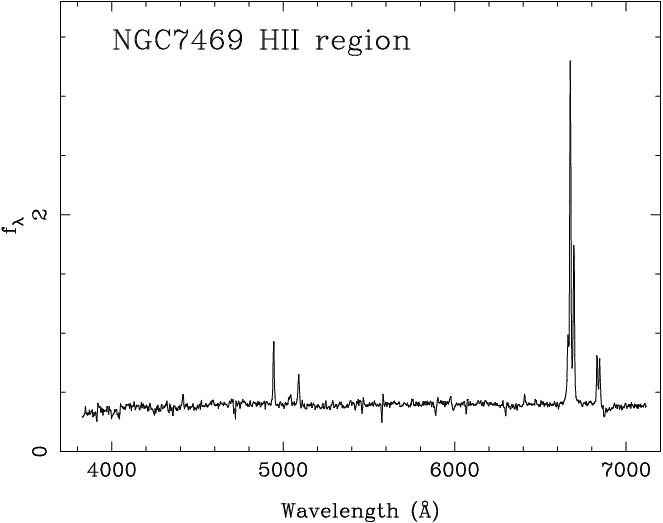}

\caption{Continued.}
\end{figure*}

\begin{figure*}
\setcounter{figure}{4}

\includegraphics[width=6.5cm,angle=-90]{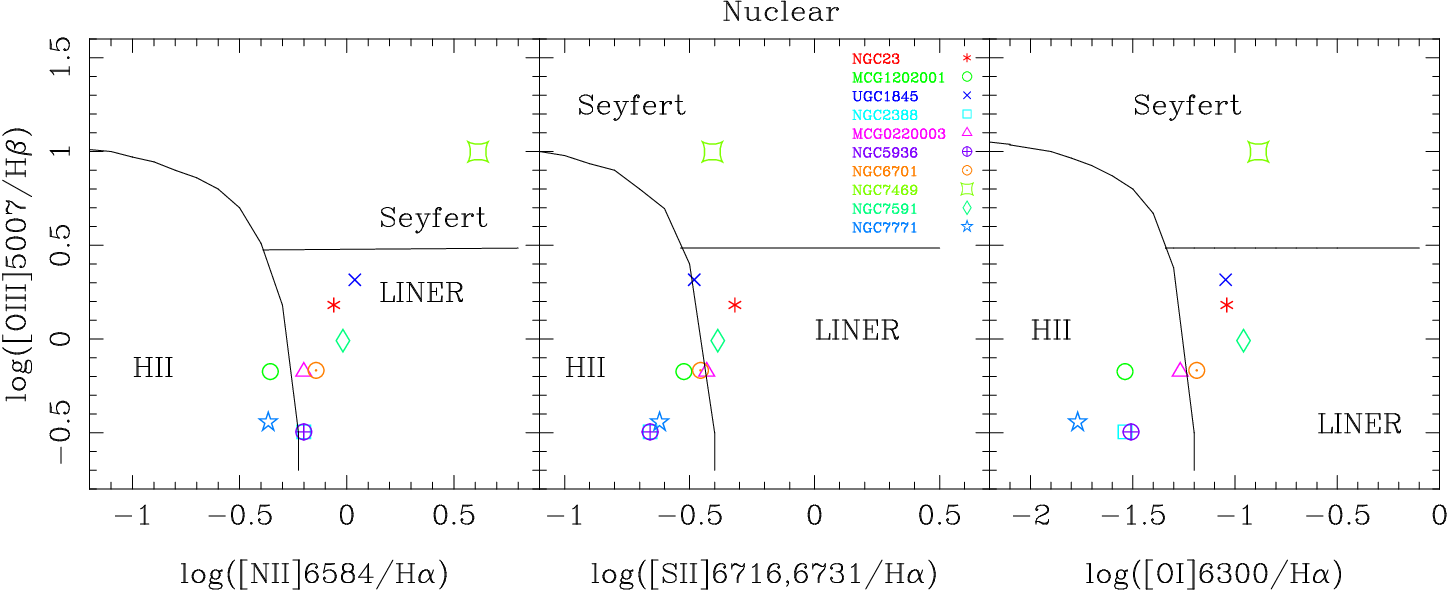}

\vspace{0.5cm}

\includegraphics[width=6.5cm,angle=-90]{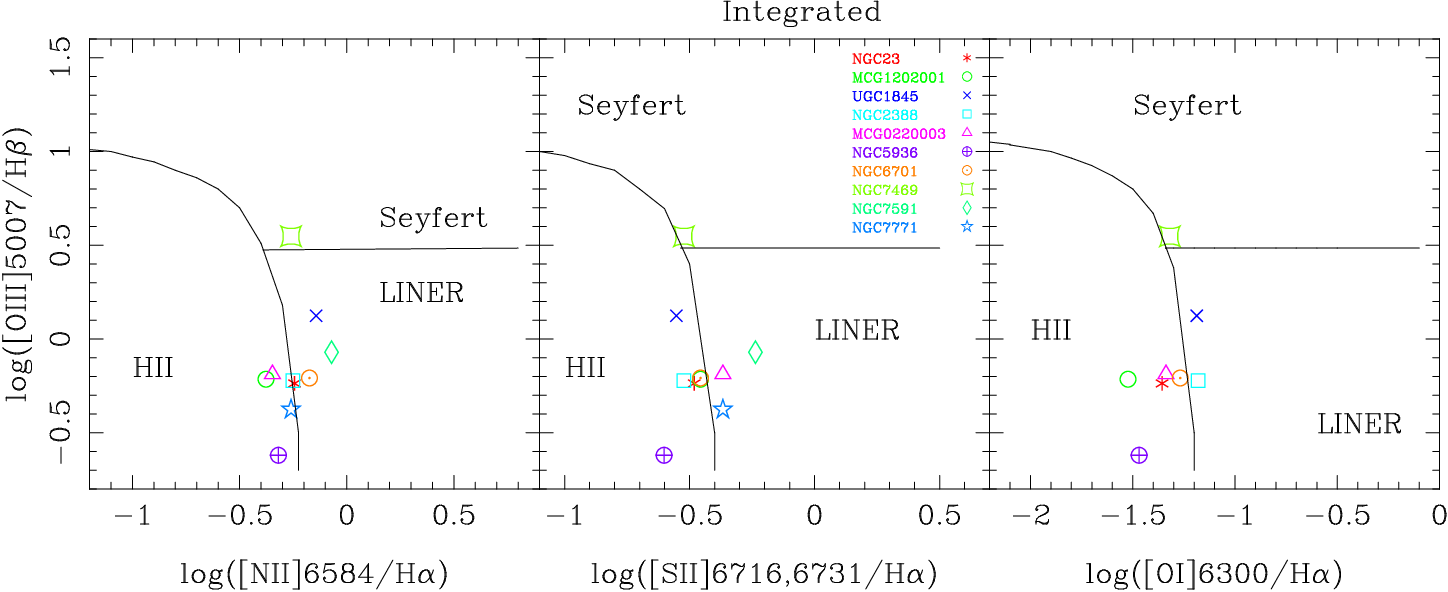}

\caption{BPT diagrams for the PMAS nuclear (upper panels) and 
the integrated (lower panels) emission of the LIRGs in our sample.  
The thin lines are the classical boundaries of 
Veilleux \& Osterbrock (1987) for the
  H\,{\sc ii} region, LINER, and Seyfert excitation. 
} 
\end{figure*} 

\begin{figure*}
\setcounter{figure}{5}

\includegraphics[width=6.5cm,angle=-90]{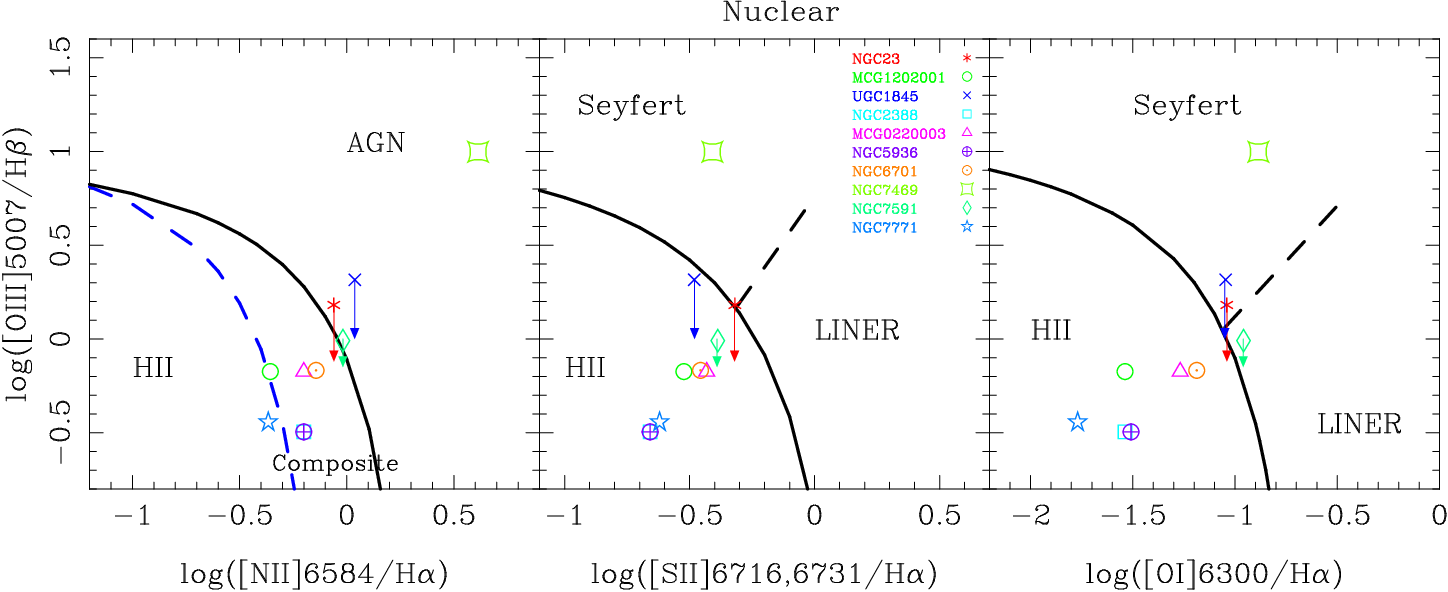}

\vspace{0.5cm}

\includegraphics[width=6.5cm,angle=-90]{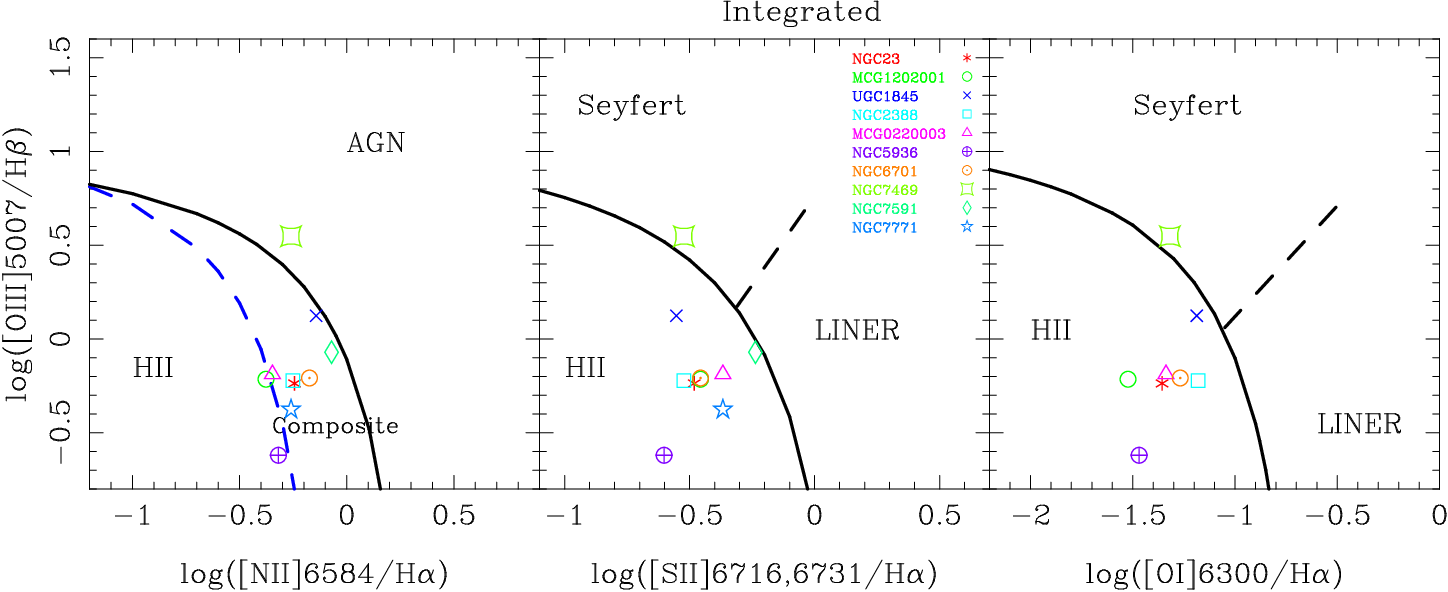}

\caption{Same as Fig.~5 but showing 
the so-called ``maximum starburst lines'' (thick solid lines),
defined by Kewley et al. (2001a) from theoretical modeling as the lines 
above which line ratios cannot be explained by star formation alone.
We also show the empirical separation between AGN and H\,{\sc
  ii} regions of Kauffmann et al. (2003), and between Seyfert and
LINER of Kewley et al. (2006), as thick dashed lines. In the upper
panel, the arrows  represent the result of correcting the H$\beta$
fluxes for underlying stellar absorption for the three
nuclei close to the AGN/HII boundaries (NGC~23, UGC~1845, and
NGC~7591).} 
\end{figure*}

\begin{table*}
\begin{center}
\caption{Spectral classifications of the nuclear and integrated spectra.}
\begin{tabular}{lcccccccc}
\hline
Galaxy        & Type        & \multicolumn{3}{c}{Classical V\&O87 Boundaries}
& \multicolumn{3}{c}{Theoretical/SSDS Boundaries} & Adopted\\
              &             & [OI] &[NII]  &[SII] & [OI] &[NII]
              &[SII]  &  \\
(1) & (2) & (3) & (4) & (5) & (6) & (7) & (8) & (9) \\
\hline
\hline
NGC~23	      & Nuclear     &L        &L        &L     & HII/L &
AGN/Composite& HII/L &  
Composite  \\
	      & Integrated  & HII      &L/HII    &HII    & HII  &
Composite & HII & HII\\
\\
MCG~+12-02-001 & Nuclear    & HII      &HII      &HII     & HII &
Composite & HII & HII \\ 
	      & Integrated  & HII      &HII      &HII & HII & Composite &
HII &HII   \\ 
\\
UGC~1845      & Nuclear     & L	&L     &L/HII &Sy/L & AGN & Sy/HII & Composite\\ 
	      & Integrated  & L/HII    &L        &HII & Sy/HII 
& AGN/Composite & HII & Composite\\
\\
NGC~2388      & Nuclear     & HII     &L/HII	  &HII & HII &
Composite & HII & HII\\
	      & Integrated  & HII     &L/HII     &HII & HII &
              Composite & HII & HII\\ 
\\
MCG~+02-20-003 & Nuclear    & L/HII   &L/HII     &L/HII& HII &
Composite & HII &  HII\\ 
	      & Integrated  & HII     &HII       &L/HII & HII &
              Composite & HII & HII\\ 
\\
IC~860	      & Nuclear     & $\cdots$       &L/Sy?      &L/Sy? &
$\cdots$ & AGN? & L? & $\cdots$  \\
\\
NGC~5939     & Nuclear     & HII     &L/HII     &HII & HII & Composite
& HII & HII \\
	      & Integrated & HII     &L/HII     &HII & HII & HII & HII
              & HII\\  
\\
NGC~6701     & Nuclear     & L       &L         &L/HII & HII/L &
Composite & HII  & Composite\\  
	      & Integrated & L/HII   &L         &L/HII & HII &
              Composite & HII & HII\\ 
\\
NGC~7469      & Nuclear     & Sy      &Sy        &Sy & Sy & AGN & Sy &
Sy\\
               & Integrated &  Sy/HII & Sy & Sy/HII & Sy/HII &
               Composite & Sy/HII & Composite\\
\\
NGC~7591      & Nuclear     & L       &L          &L & L &
AGN/Composite & HII/L & Composite\\      
	      & Integrated  & $\cdots$ & L/HII & L/HII & $\cdots$ & 
AGN/Composite & L/HII & Composite\\
\\
NGC~7771     & Nuclear     & HII     &HII        &HII & HII & HII &
HII & HII\\
	      & Integrated  & $\cdots$        &L/HII      &L/HII & 
$\cdots$ & Composite & HII & HII\\ 

\hline
\end{tabular} 
\end{center}
Notes.--- Column~(2): PMAS type of 1D spectra.
Columns~(3), (4), and (5):
Spectral classification based on the [O\,{\sc
  iii}]$\lambda$5007/H$\beta$ vs.   
[O\,{\sc i}]$\lambda$6300/H$\alpha$, [N\,{\sc
  ii}]$\lambda$6584/H$\alpha$, 
and 
the [S\,{\sc ii}]$\lambda\lambda$6717,6731/H$\alpha$ diagrams,
respectively and the classical
boundaries of V\&O87 in the BPT diagrams. 
The classifications are, HII=HII galaxy, L=LINER, and Sy=Seyfert. 
Galaxies classified as Composite are likely to be a combination
of AGN  activity and star formation. The classifications take into
account the $\pm 1\sigma$ uncertainties reported by Kewley et
al. (2001a) for the maximum starburst lines.
Columns~(6), (7), and (8):
As Columns~(3), (4) and (5) but using the latest theoretical/SSDS
boundaries. 
Column~(9): Adopted spectral classification.\\

\end{table*}

We did not attempt to correct the line ratios for extinction. The effects of extinction on the spectral
classifications should be moderately small because
the lines involved in the line ratios are in most cases close in
wavelength. 

The first result worth noticing is that when using the  
[O\,{\sc iii}]$\lambda$5007/H$\beta$ vs. 
[N\,{\sc ii}]$\lambda$6584/H$\alpha$  
diagram and the theoretical/SSDS boundaries (Fig.~6 and Table~3), 
a large fraction of the LIRG nuclei are classified as composite, and
thus they are likely to contain a metal-rich stellar population and an 
AGN (see Kewley et
al. 2006). It is also possible that some of these composite objects 
have an added contribution
from shock excited emission (from supernovae) associated with an 
aging starburst (see Alonso-Herrero et al. 2000b). If we use all three
BPT diagrams the fraction of composite 
objects appears to be smaller. However, Kauffmann et al. (2003) did
not provide an empirical separation between AGN and star-forming  
 galaxies in the diagrams involving the [O\,{\sc i}]$\lambda$6300/H$\alpha$ and
 the [S\,{\sc ii}]$\lambda\lambda$6717,6731/H$\alpha$ ratios, so we
 cannot assess whether galaxies are composite or not using those two diagrams.

Using all three BPT diagrams and 
the V\&O87 classical boundaries (thin
lines in Fig.~5) we find
that only  two nuclei in our sample of LIRGs show pure H\,{\sc ii}-like
excitation (that is, classified as H\,{\sc ii} using all three BPT
diagrams), five  have an intermediate LINER/H\,{\sc ii}
classification, two fall in 
the LINER regions, and one nucleus is classified as a Seyfert galaxy
(upper panels of Fig.~5, see also Table~3).
This large fraction of composite objects (i.e., star formation and AGN
activity)   is a well-known property of samples of
infrared-bright galaxies (Veilleux et al. 1995; Kewley et al. 2001b;
Chen et al. 2009), regardless of what boundaries are used for
  classifying galaxies.

 We
cannot provide a conclusive classification for IC~860 because the
emission lines in the blue part of the spectrum are not detected (note
the strong H$\beta$ absorption, Fig.~4). 
The high [N\,{\sc ii}]$\lambda$6584/H$\alpha$ and 
[S\,{\sc ii}]$\lambda\lambda$6717,6731/H$\alpha$ line
ratios could indicate a Seyfert or LINER classification, 
  but the correction for underlying stellar H$\alpha$ absorption would
  decrease significantly the observed line ratios in this galaxy.

A large fraction of the LIRGs in this sample show a strong
  contribution from an evolved stellar population as indicated by the
  presence of absorption features (Balmer line series, H+K CaII
    lines) in the blue part of the spectra
  (Fig.~4). Although  a detailed modeling of the stellar populations
is beyond the scope of this paper, we can attempt to correct  
the [O\,{\sc  iii}]$\lambda$5007/H$\beta$  ratio for the presence of
an evolved stellar population for those galaxies located near the
AGN/HII boundaries. These are NGC~23, UGC~1845, and NGC~7591
(Fig.~6). Our preliminary modeling indicates
that a combination of intermediate ($1-5\,$Gyr) 
and young ($<10\,$Myr) stellar  populations with
a measured equivalent width of H$\beta$ in absorption of 
$\sim 4$\AA\,
would produce acceptable fits to their spectra. This is within the
average H$\beta$ stellar absorption corrections obtained by Moustakas \&
Kennicutt (2006) for a sample of nearby star-forming galaxies. 
The approximate effects of correcting the observed 
[O\,{\sc  iii}]$\lambda$5007/H$\beta$ for underlying stellar
absorption are shown as arrows in the upper panel of Fig.~6. The
corrections for the other line ratios involving H$\alpha$ would be
smaller. As can be
seen from this figure, this correction does not change fundamentally
the result that these nuclei appear to be composite in nature.

\begin{table}
\begin{center}
\caption{Gas kinematic results.}
\begin{tabular}{lccc}
\hline
Galaxy        & $cz_{{\rm nuc, H}\alpha}$ &
  $\Delta v_{{\rm H}\alpha}$       & 
$\sigma_{{\rm nuc, H}\alpha}$\\
(1) & (2) & (3) & (4) \\
\hline
\hline
NGC~23	       & $4495\pm 9$ & $-215$, $+236$ & $139\pm 17$ \\
MCG~+12-02-001 & $4732\pm 8$ & $-110$, $+156$ & $105\pm 14$\\
UGC~1845       & $4796\pm 9$ & $-237$, $+181$ & $142\pm 17$\\
NGC~2388       & $4093\pm 9$ & $-157$, $+163$ & $115\pm 15$\\
MCG~+02-20-003 & $4950\pm 9$ & $-141$, $+110$ & $96\pm 13$\\
IC~860$^*$     & $4146\pm 13$ & $\cdots$      & $\cdots$  \\
NGC~5936       & $3984\pm 13$ & $-66$, $+179$ & $66\pm 15$\\
NGC~6701       & $3946\pm 13$ & $-41$, $+102$ & $92\pm 16$\\  
NGC~7469       & $4946\pm 9^{**}$  & $-110$,$+207$$^{**}$ & $125\pm 16^{**}$ \\
NGC~7591       & $4869\pm 9$ & $-179$, $+345$  & $144 \pm 17$\\     
NGC~7771       & $4334\pm 9$ & $-181$, $+199$ & $106\pm 14$\\
\hline
\end{tabular} 
\end{center}

Notes.---  Column~(2): $cz_{{\rm nuc, H}\alpha}$
is the H$\alpha$ velocity of the 
  nucleus, defined as the peak of the 6200\,\AA \ continuum, except
  for NGC~7771 where it is refered to the position that makes the
  velocity field gradient symmetric (see text and Fig.~3c). 
Column~(3): $\Delta v_{{\rm H}\alpha}$ is the H$\alpha$
peak-to-peak velocity range over the PMAS FoV. Column~(4): 
$\sigma_{{\rm nuc, H}\alpha}$ is the nuclear H$\alpha$ velocity dispersion
(corrected for instrumental effects). All
  measurements are  in km s$^{-1}$, and are not corrected for
  inclination and/or angular resolution effects.\\
$^*$For IC~860 $cz_{\rm nuc}$ is from  [N\,{\sc ii}]$\lambda$6584 and 
[S\,{\sc ii}]$\lambda\lambda$6717,6731. We note that
NED quotes a value of $3347\,{\rm
  km\,s}^{-1}$ from 21\,cm H\,{\sc i} measurements, whereas our value
of the velocity is confirmed by measurements of the mid-infrared emission
lines with {\it Spitzer}/IRS (Pereira-Santaella et al. 2009, in
preparation).\\ 
$^{**}$For NGC~7469 the measurements are from 
[N\,{\sc ii}]$\lambda$6584.
\end{table}

We have seven  galaxies in common with the work of Veilleux et al. (1995). Our
nuclear classifications 
are in good agreement with theirs. The only two exceptions are 
NGC~23  and NGC~6701, which we would classify as
LINER/HII with the V\&O87 boundaries, or as composite using the 
theoretical/SSDS boundaries. As explained in Sect.~3.2 we extracted
our nuclear spectra with 
the smallest possible physical sizes allowed by the PMAS spaxels 
($\sim 300\,$pc, Table~2), whereas 
Veilleux et al. (1995) used linear sizes of 2\,kpc for 
extracting their nuclear spectra. As can be
seen from Fig.~1a and 1h, the Veilleux et al. (1995) apertures 
included a large number of H\,{\sc ii} regions in the ring of star
formation of NGC~23 and in the inner spiral structure of NGC~6701. This
readily explains the H\,{\sc ii}-like classification given by Veilleux
et al. (1995).

The BPT diagrams for the integrated emission over the PMAS FoV are
presented in the lower panels of Figs.~5 and 6. 
The integrated line ratios of 
the four galaxies whose nuclei 
are classified as LINERs using the V\&O87 boundaries 
now fall in the H\,{\sc ii} region or in the
intermediate LINER/H\,{\sc ii} region. A similar situation is
  seen for NGC~7469 for which the integrated line ratios move toward
  the composite area in all diagrams. This is well understood in
terms of the increased contribution of extra-nuclear high surface-brightness 
H\,{\sc ii} regions (see H$\alpha$
morphologies in Fig.~1) to their integrated emission. For the other
galaxies there is no general trend, as the integrated line
  ratios depend on the
relative contribution of H\,{\sc ii} regions and diffuse emission to
the total line
emission over the PMAS FoV (Alonso-Herrero et al. 2009, in
preparation). For instance, the integrated line ratios of NGC~7771 
show larger [N\,{\sc ii}]$\lambda$6584/H$\alpha$ and 
[S\,{\sc ii}]$\lambda\lambda$6717,6731/H$\alpha$ line ratios 
than the nuclear ones,
whereas the nuclear and integrated 
line ratios of other galaxies (e.g., MCG+02-20-003) 
remain approximately constant or slightly more similar to those of  H\,{\sc
  ii} regions. 

Finally, we show an example of the power of optical IFS in identifying 
different excitation conditions. For NGC~7469  we also
extracted the spectrum of an H\,{\sc ii} region in
the circumnuclear ring of star formation, located at about 2\arcsec \,
west from
the nucleus (Fig.~4). It is clear that the line ratios are typical of H\,{\sc
  ii}-like excitation and are not contaminated by the nearby Seyfert 1
nucleus (i.e., no broad components are present in the hydrogen
recombination lines). 

Summarizing, the 
comparison of the nuclear and integrated activity classifications,
together with the spatial distribution of the bright emission lines
(in particular H$\alpha$) allowed us to isolate and quantify the different
ionization sources
(nuclei, and circumnuclear H\,{\sc ii} regions and diffuse emission)
contributing to the observed emission in galaxies (see also
Alonso-Herrero et al. 2009, in preparation).

\section{Ionized Gas Kinematics}

The velocity fields of the emission lines over the
central few kpc (typically 5\,kpc) are mostly consistent with  rotation 
(see Fig.~3) for all the galaxies in the sample. We note  that even 
on the physical scales probed by PMAS ($\sim 300\,$pc), 
some velocity fields appear to show some peculiarities. 
However, it is clear that these velocity fields are more similar to
those of disk galaxies 
(Falc\'on-Barroso et al. 2006: Daigle et
al. 2006) than to those of local ULIRGs 
(Colina et al. 2005: Monreal-Ibero et al. 2006).   
The
large scale continuum major photometric axis and the H$\alpha$ major kinematic 
axis are broadly in agreement with each other (seen in
projection) in the majority of our LIRGs.
 The ionized gas peak-to-peak velocities of LIRGs 
(over the central $\sim 5\,$kpc)
 are  typically between $200$ and $400\,{\rm km\,s}^{-1}$  (see
 Table~4), whereas in ULIRGs with tidally induced flows the
gas peak-to-peak velocities can be as high as $600\,{\rm km\,s}^{-1}$
(Colina et al. 2005).

Three galaxies in our sample show circumnuclear rings of star
formation. The central H$\alpha$ velocity field of NGC~23 
is consistent with
rotation, with the major kinematic axis aligned with the major axis of
the galaxy rather than with the orientation of the ring of star formation
as seen in  H$\alpha$  (Fig.~1a).
A similar situation is observed for 
NGC~7469, where the [N\,{\sc ii}]$\lambda$6584 velocity field has a
major axis similar to that of the continuum, although it is 
asymmetric towards the north-west direction. This asymmetry was
already reported in the rotation curve along the major axis of the
galaxy by M\'arquez \& Moles (1994). In both galaxies, it 
is likely that 
most of the mass in the central region is in relatively evolved stars
rather than in young ionizing stars (see also D\'{\i}az-Santos et
al. 2007 for NGC~7469). 
The H$\alpha$ velocity field of the inner  $\sim 10\arcsec$ 
of NGC~7771 has a major kinematic axis aligned with the approximate
orientation of the circumnuclear ring of star formation (east-west
direction), whereas the outer velocity 
field appears to have a major kinematic axis in better agreement with that
of the large scale continuum emission. This kind of velocity fields with
symmetric distortions can be associated with the presence of a warped
disk. It is also worth mentioning that there is  dynamical evidence
that NGC~7771 is weakly
interacting with NGC~7770 (Keel 1993) and is located in a group of galaxies. 

Another example of a central H$\alpha$ velocity field that deviates from
perfect rotation is that
of NGC~6701. This galaxy shows a complex overall 
morphology, with the presence of a large
scale bar, an inner isophote twist produced by a spiral like inner ring
(see M\'arquez et al. 1996, and the continuum $1.6\,\mu$m image in
Fig.~1h), a perturbed rotation curve, 
and a small companion likely to be responsible for some of these
properties (M\'arquez et al. 1996). As seen in other galaxies in our sample, the
major kinematic axis of the H$\alpha$ velocity field appears to be better 
aligned with the photometric axis of
the galaxy (PA$=114\degr$, Vogt et al. 2004)  
rather than with the orientation of the nuclear high 
surface-brightness emission Pa$\alpha$ emission, which is seen in an
almost north-south 
orientation. 

Arp~299 the most
luminous galaxy in our sample and a
strongly interacting system shows very complicated
velocity fields (both of neutral and ionized gas,
see Garc\'{\i}a-Mar\'{\i}n et al. 2006), not only at the
interface between the two galaxies, but also in the nuclear regions of
the two members of the system. Moreover, the velocity fields of the
ionized gas in Arp~299 do 
not appear to be dominated by ordered virialized motions, probably 
as a result of the interaction between the two galaxies, as is 
the case of most ULIRGs.

As can be seen from Fig.~3 for most LIRGs in our sample the 
peak of the H$\alpha$ velocity
dispersion coincides with the peak of the optical and near-infrared
continuum emission (i.e., the nucleus), and thus the velocity
dispersion is likely to be tracing mass. The nuclear H$\alpha$
velocity dispersions are between 66 and 144\,${\rm km \,s}^{-1}$
(Table~4). The largest velocity dispersions 
of the ionized gas are associated with three of the nuclei classified
as LINERs and the Seyfert galaxy NGC~7469. 

In the case of galaxies with circumnuclear rings of star formation, we
find that the gas velocity dispersions in the rings 
are less than in the nuclear
regions. This is consistent with the findings of Falc\'on-Barroso et
al. (2006) for the same type of galaxies and was interpreted as an
indication for the presence of large amounts of cold gas from which
stars have recently formed. 

There are only a few measurements of the stellar velocity 
dispersion of LIRGs. From near-infrared CO
absorption features the stellar velocity dispersions are between 60 and 
160\,${\rm km \,s}^{-1}$ (Shier, Rieke, \& Rieke 1996; Hinz \& Rieke
2006), similar to the range of H$\alpha$ velocity dispersions measured
for our LIRGs. 
The only LIRG in our sample with a stellar velocity dispersion
measurement is NGC~7469, for which Onken et al. (2004) obtained 
$\sigma_*=142 \pm 16\,{\rm km\,s}^{-1}$ from the calcium triplet, in
good
agreement with our measurement from the ionized gas. 
The velocity dispersions of LIRGs are in 
general similar to or slightly less than the typical nuclear gas
and stellar velocity dispersions 
 of ULIRGs   (Colina et al. 2005; Dasyra et al. 2006). This suggests 
that at least some LIRGs  have comparable  
dynamical masses to those of ULIRGs, although a detailed modeling is
 needed to assess this issue.

\section{Discussion and Summary}
This is the first paper in a series presenting PMAS optical IFS
observations of the northern hemisphere portion of the volume-limited 
($v=2750-5200\,{\rm km\,s}^{-1}$) 
sample of local LIRGs defined by Alonso-Herrero et al. (2006). This
sample is in turn  
part of the larger IFS survey of nearby ($z<0.26 $) LIRGs and ULIRGs 
assembled by Arribas et
al. (2008).  In this paper
we presented the observations and 
data reduction of the PMAS observations. The PMAS observations cover 
the central $16\arcsec \times 16\arcsec$ (typically the central 
5\,kpc) with spaxels of 1\arcsec \, in size, and a spectral range 
$\sim 3800-7200\,$\AA. The PMAS IFS data were complemented
with our own existing near-infrared {\it HST}/NICMOS observations of
the $1.6\,\mu$m 
continuum and the Pa$\alpha$
emission line. The main goal of this paper is to present
an atlas of the observations and 
the general IFS results of the sample of LIRGs
and compare them with local ULIRGs.

On the physical scales probed by the PMAS IFS ($\sim
300\,$pc) the optical and near-infrared stellar morphologies
are similar for most galaxies, indicating that extinction is not
playing a major role for this sample of LIRGs, except in the innermost regions.  Similarly, 
there are no major morphological differences between 
Pa$\alpha$ and H$\alpha$. The {\it
  HST}/NICMOS Pa$\alpha$ and PMAS H$\alpha$ observations are
complementary, with the former 
revealing in great detail (physical scales of a few 
tens of parsecs) the morphologies of the
high surface-brightness H\,{\sc ii} regions, and the latter 
being  sensitive not only to H\,{\sc ii} regions but also to 
diffuse emission with lower surface brightness. 

In the majority of the LIRGs in our sample 
 the peaks of the continuum and gas (e.g., H$\alpha$, [N\,{\sc
   ii}]$\lambda$6584) emissions coincide. This contrasts with  
local interacting ULIRGs, where the extreme effects of the interaction
processes on the ionizing mechanisms and the distribution of dust 
can  
cause displacements between the peaks of continuum and gas emission of
 typically $2-4\,$kpc (Colina et al. 2005: Garc\'{\i}a-Mar\'{\i}n et
al. 2009). The only exceptions in the LIRG sample are galaxies with
circumnuclear rings of star formation (NGC~23 and NGC~7771) 
and the strongly interacting galaxy Arp~299. In the case of the
galaxies with circumnuclear rings of star formation, the most luminous
H$\alpha$ emitting regions are found in the rings rather
than in the nuclei of the galaxies, and the displacements are well
understood in terms of differences in the 
stellar populations. In Arp~299 the displacements are due to the high 
extinctions suffered by the nuclear regions 
(see Alonso-Herrero et al. 2000a and 
Garc\'{\i}a-Mar\'{\i}n et al. 2006).

Using standard BPT diagrams we compared the excitation conditions of
the nuclear and 
integrated (over the PMAS FoV) 1D spectra of the LIRGs in our
sample. Only two nuclei show
 pure H\,{\sc ii}-like excitation using the classical 
V\&O87 boundaries, and one
has a Seyfert nucleus. 
The rest are classified as LINER or intermediate LINER/H\,{\sc ii},
using the V\&O87 boundaries, or alternatively as composite 
objects using the
  theoretical/SSDS boundaries.  We also found that a large
  fraction of the nuclei  show evidence
for a strong contribution from an 
evolved stellar population (absorption features). 
There is no general trend for the  excitation
conditions of the integrated emission when compared with  the nuclear
excitation, as the former depends on the
relative contributions of H\,{\sc ii} regions and the diffuse emission
to the line emission over the PMAS FoV (Alonso-Herrero et al. 2009, in
preparation). That is, galaxies dominated by high surface-brightness
H\,{\sc ii} regions show integrated H\,{\sc ii}-like excitation,
whereas galaxies with more diffuse low surface-brightness   
emission tend to show 
  slightly larger [N\,{\sc ii}]$\lambda$6584/H$\alpha$ and 
[S\,{\sc ii}]$\lambda\lambda$6717,6731/H$\alpha$
line ratios.

The H$\alpha$ velocity fields over the central few kpc covered by the
PMAS observations are generally consistent, at least to first order, with
rotational motions.
The observed H$\alpha$ velocity amplitudes (peak-to-peak) are between 200 and 
$\sim 400\,{\rm
  km\,s}^{-1}$. Although the velocity fields of some LIRGs show some
peculiarities, they are not as perturbed as those of  most local,
strongly interacting ULIRGs.  In that respect, the velocity fields of
the emission lines of our sample resemble those of disk galaxies
(Falc\'on-Barroso et al. 2006: Daigle et al. 2006).
In most LIRGs in our sample the peak of the H$\alpha$ velocity
dispersion coincides with the peak of the optical and near-infrared
continuum emission (i.e., the nucleus). Thus, the velocity
dispersion is likely to be tracing mass and provides further support
for rotation. The nuclear H$\alpha$
velocity dispersions are in the range 
$\sigma_{{\rm nuc,H}\alpha}=66 - 144\,{\rm km \,s}^{-1}$ and
are similar to the stellar values measured from 
near-infrared CO absorption features measured for other LIRGs. 
The LIRG nuclei with the largest H$\alpha$ velocity dispersions are
those classified as LINERs and the Seyfert 1 nucleus of NGC~7469.

Throughout this paper we discussed the ionized gas and stellar
distributions, excitation conditions and kinematics of a volume
limited sample of LIRGs. We also showed that the properties of this
sample of LIRGs, and in particular, their kinematics, 
 are more similar to those of disk galaxies, rather than to
those of local, strongly interacting  ULIRGs. 
Our LIRGs are part of a flux and volume limited sample 
drawn from the {\it IRAS} RBGS of 
Sanders et al. (2003). As a consequence,  our 
sample is predominantly composed of 
low-luminosity LIRGs with an average IR luminosity of 
 $\log (L_{\rm IR}/{\rm L}_\odot) \sim 11.32$
for the full sample (northern and southern hemispheres, see
Alonso-Herrero et al. 2006). 
Sanders \& Ishida (2004) demonstrated that 
$\log (L_{\rm IR}/{\rm L}_\odot) \sim 11.50$  marks the 
transition between samples being dominated by
disk galaxies and merger dominated samples. In the 
Sanders \& Ishida (2004) sample of LIRGs, most
objects at  $\log (L_{\rm IR}/{\rm L}_\odot) < 11.40$
are spiral galaxies with no signatures of a
major interaction and pairs of galaxies, whereas at 
$\log (L_{\rm IR}/{\rm L}_\odot) > 11.70$ most objects are
strongly interacting equal mass galaxies with overlapping disks. 
The similarities of our sample with  ¨normal¨ disk
galaxies are then well understood because our sample 
is mostly composed of disk galaxies and galaxy pairs not undergoing a
strong interaction (Sect.~2.1). 

At intermediate redshifts ($z \sim 1$) LIRGs are the main contributors to the
star formation rate density 
(Elbaz et al. 2002; Le Floc'h et al. 2005; P\'erez-Gonz\'alez et al. 2005; 
Caputi et al. 2007). Moreover, these $z\sim 1$ LIRGs 
are mostly classified as spiral galaxies
(Bell et al. 2005: Melbourne et al. 2005, 2008) 
with the star formation smoothly distributed in the disks of the
galaxies as seen for our sample of LIRGs. There is also a strong 
evolution in galaxy kinematics at $z<1$, but still about half of 
the line emitting population at these redshifts, which are mostly
  LIRGs,  have rotating disks (Puech et al. 2008; Yan et
al. 2008).
While local and  distant LIRGs may have self regulated star formation as in disk
galaxies (see Bell et al. 2005), it is clear that their star formation 
rates are at least a factor of ten or
more higher than in spirals. 
Locally the central regions of LIRGs are found to contain a large
population of giant HII regions not observed in normal spiral galaxies
(see Alonso-Herrero et al. 2006).  This, together with a higher
star formation efficiency of the dense gas in local LIRGs (Graci\'a-
Carpio et al. 2008) may explain
the  high star formation rates of LIRGs when compared to normal disk
galaxies.
Studying the spatially resolved properties of flux-limited
complete samples of local LIRGs may help us understand if the
similarities with $z \la 1$ LIRGs stem from the same physical
processes and comparable evolutionary states.

Summarizing, we 
demonstrated the  ability of optical IFS to spatially resolve 
the different ionization sources  contributing to the observed
emission of LIRGs, as well as to study their kinematic properties.
Full detailed studies of the extinction, excitation conditions, 
stellar populations, and kinematics of LIRGs
and ULIRGs will be presented in forthcoming papers. 
\, 

\begin{acknowledgement}
We are grateful to the Calar Alto staff, and in particular to S. S\'anchez, 
A. Guijarro, L. Montoya, and N. Cardiel, for their
support during the PMAS observing campaigns. We thank
  Martin Roth for interesting and useful discussions about the PMAS
  instrument, and J. Rodr\'{\i}guez Zaur\'{\i}n for advice on stellar
  populations. We thank the referee for his/her constructive comments.

This research has made use of the NASA/IPAC Extragalactic Database
(NED) which is operated by the Jet Propulsion Laboratory, California
Institute of Technology, under contract with the National Aeronautics
and Space Administration.

AA-H, LC, SA, AL, and JA-G acknowledge support
from the Spanish Plan Nacional del Espacio under grants ESP2005-01480
and ESP2007-65475-C02-01. 
AA-H also acknowledges support for this work from the
  Spanish Ministry of Science and Innovation through Proyecto Intramural
Especial under grant number 200850I003. 
MG-M is supported by the German Federal
Department of Education and Research (BMBF) under project numbers: 
50OS0502 and 50OS0801. AM-I is supported
by the Spanish Ministry of Science and Innovation (MICINN) under program
"Specialization in International Organisms", ref. ES2006-0003.
\end{acknowledgement}

\clearpage

\begin{figure*}[h]
\setcounter{figure}{0}

\includegraphics[width=14.cm]{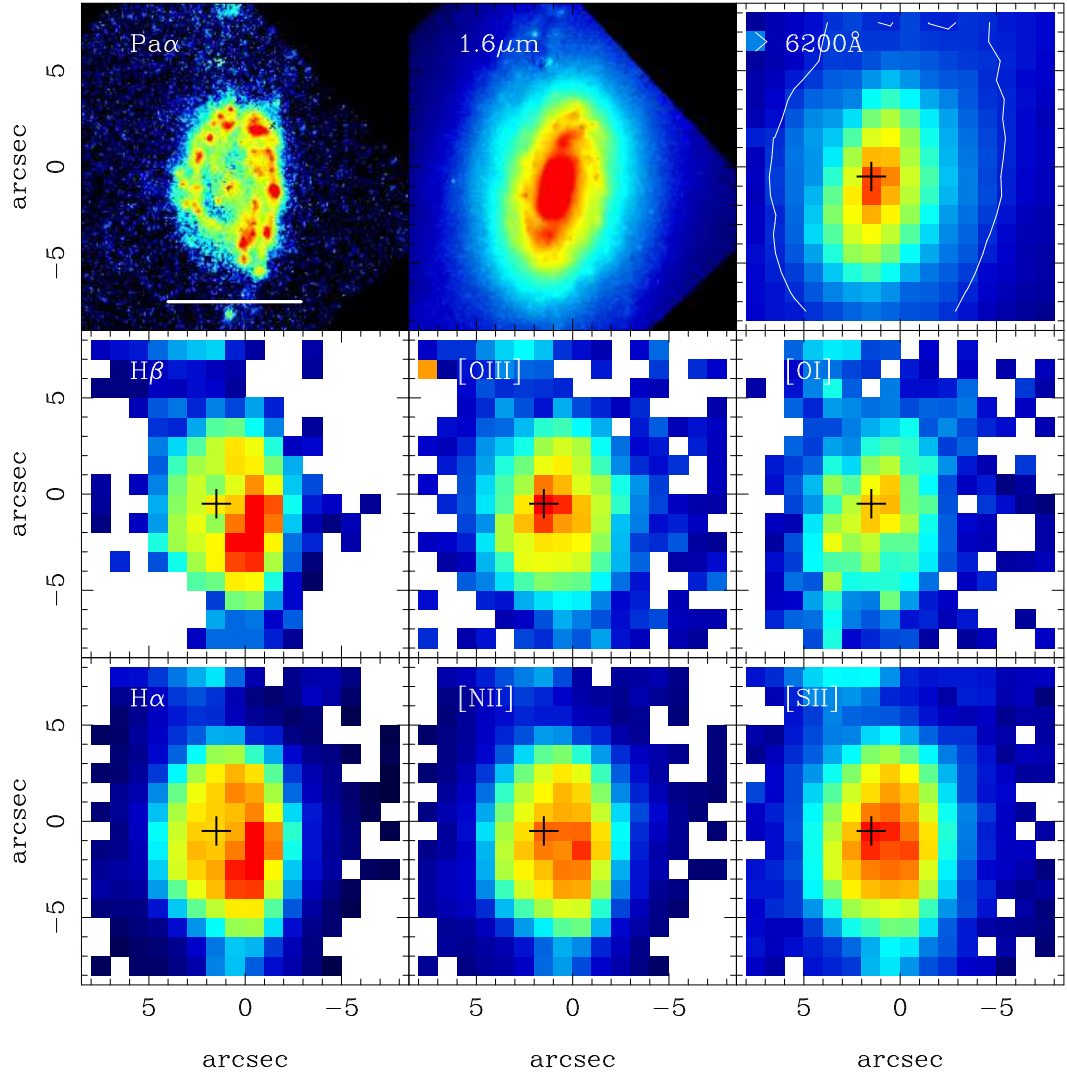}

\caption{(a) NGC~23. The middle and bottom panels are the 
PMAS observed (not corrected for extinction) 
maps of the brightest emission lines: H$\beta$, [O\,{\sc iii}]$\lambda$5007,
[O\,{\sc i}]$\lambda$6300, H$\alpha$, [N\,{\sc ii}]$\lambda$6584, and the sum of
the [S\,{\sc ii}]$\lambda\lambda$6717, 6731 lines.
The H$\beta$ map has not been corrected for stellar absorption.
The upper left panel is the map of the {\it HST}/NICMOS Pa$\alpha$ 
emission line. The maps of the PMAS 6200\,\AA \, 
and the  {\it HST}/NICMOS $1.6\,\mu$m continuum emission are  
the upper right and the upper middle panels, respectively, both
representing the stellar emission. The cross on the 
PMAS maps shows the location of the PMAS 6200\,\AA \, continuum
peak. Due to the lack of absolute astrometry of the PMAS observations,
the PMAS continuum peak is not shown on the NICMOS images. 
The horizontal bar in the Pa$\alpha$ panel represents the 
2\,kpc linear scale used by Kim et al. (1995) and Veilleux
et al. (1995) for extracting their nuclear spectra.  
The contour shown on the PMAS continuum map corresponds to the external
isophote used for extracting the integrated 1D spectra (see Sect.~3.2).
The orientation of the images is north up, east to the left. 
All the images are shown in a square root scale.
 }
\end{figure*}

\begin{figure*}
\setcounter{figure}{0}
\includegraphics[width=14.cm]{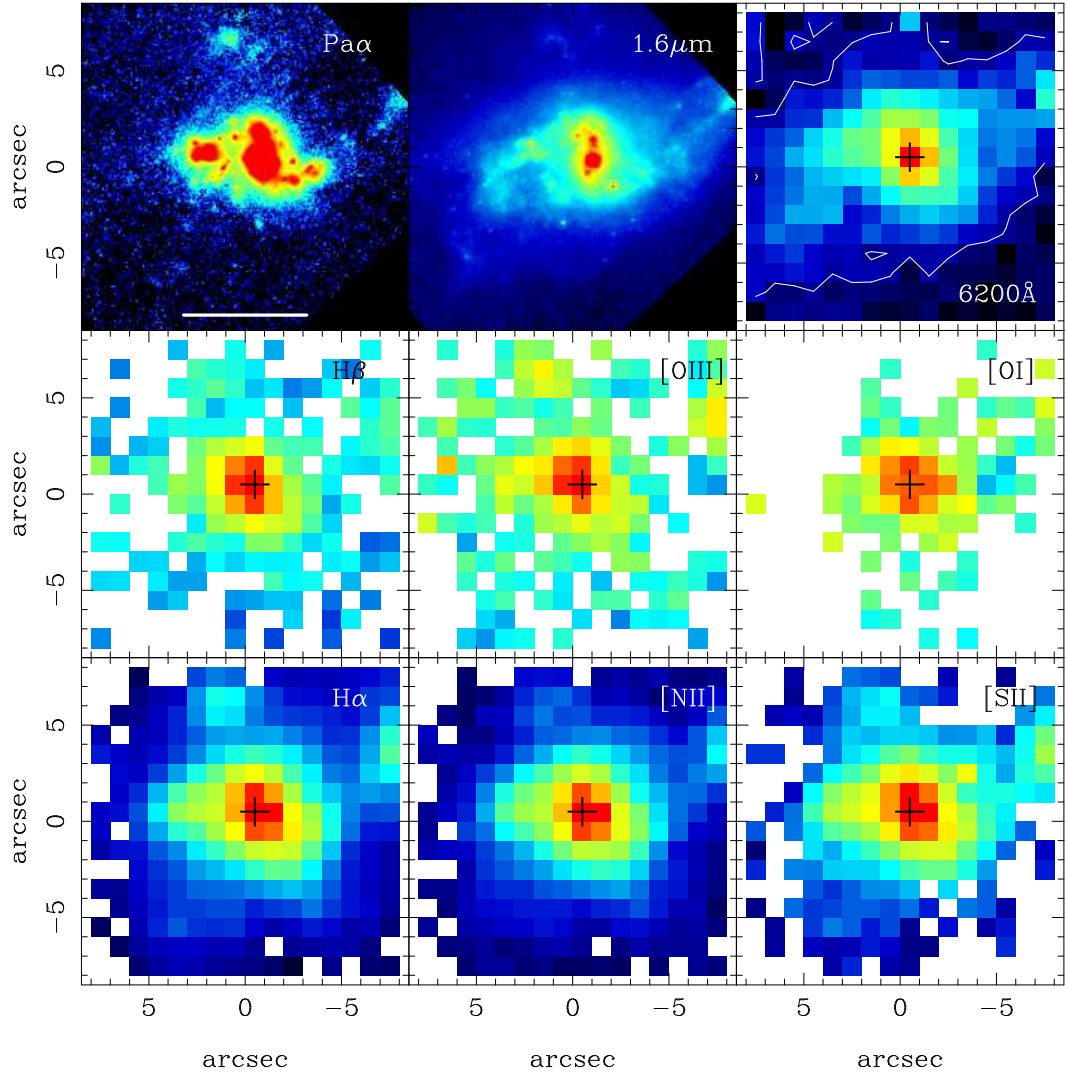}
\caption{(b) As Fig.~1a but for MCG~+12-02-001.}
\end{figure*}

\begin{figure*}
\setcounter{figure}{0}
\includegraphics[width=14.cm]{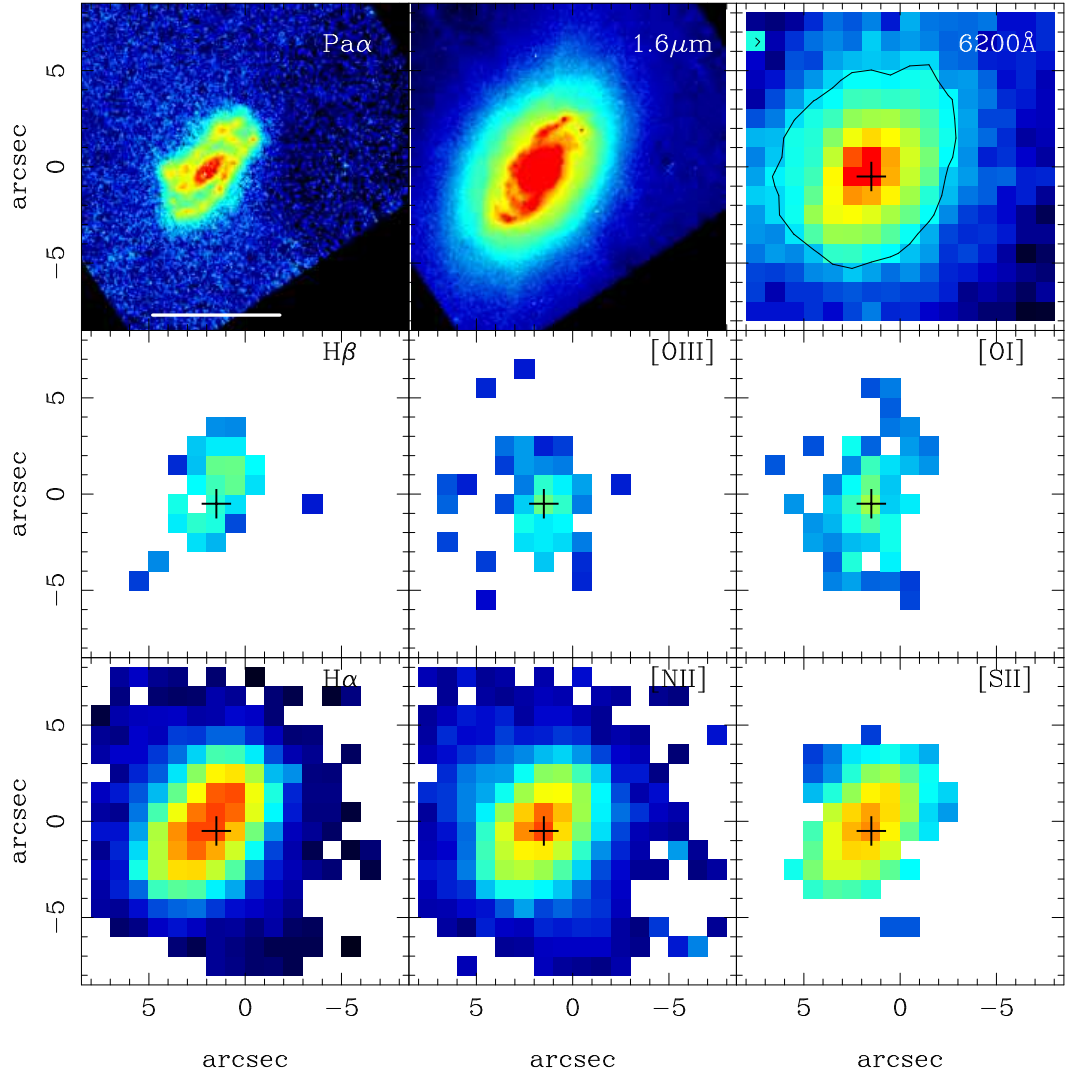}
\caption{(c) As Fig.~1a but for UGC~1845.}
\end{figure*}

\begin{figure*}
\setcounter{figure}{0}
\includegraphics[width=14.cm]{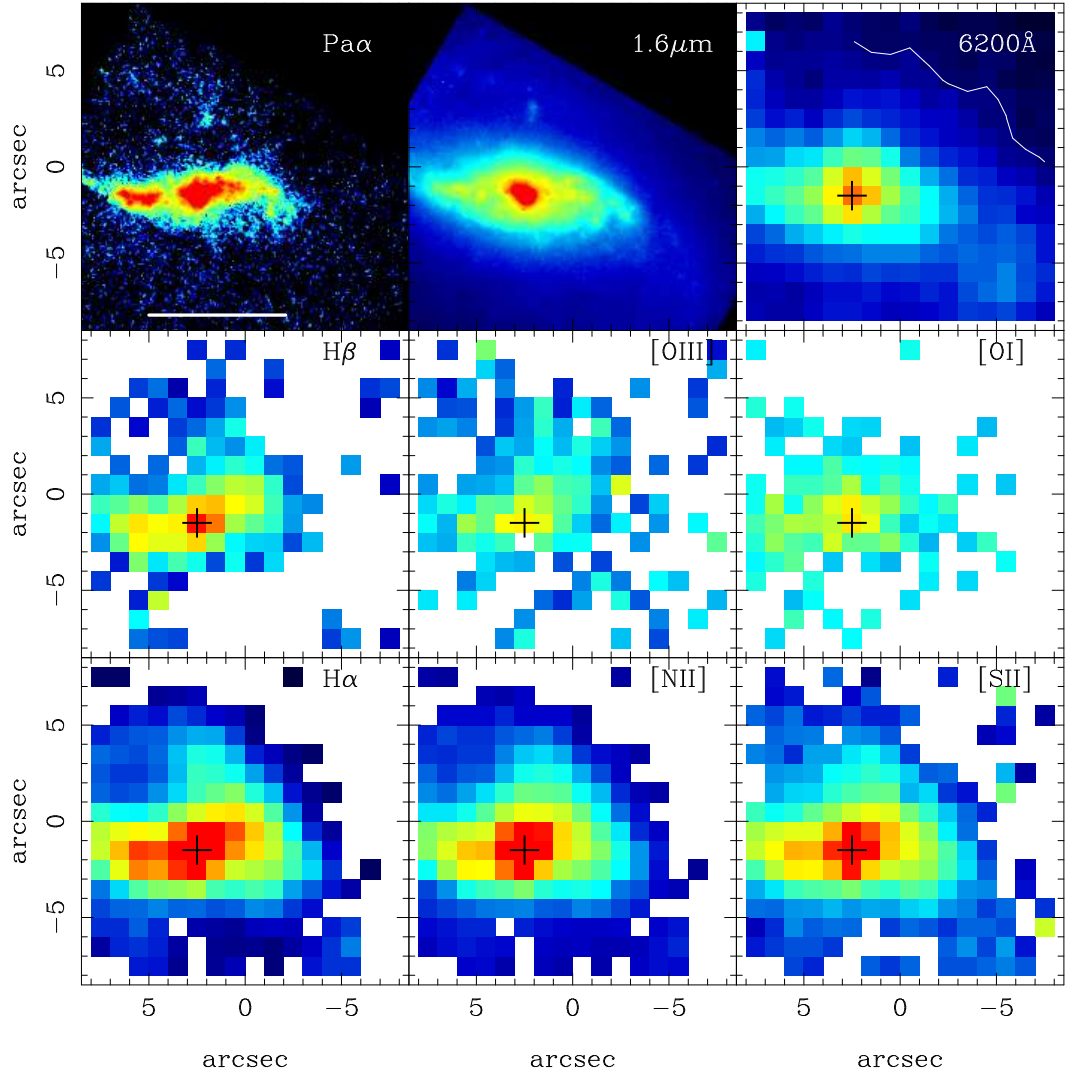}
\caption{(d) As Fig.~1a but for NGC~2388.}
\end{figure*} 

\begin{figure*}
\setcounter{figure}{0}
\includegraphics[width=14.cm]{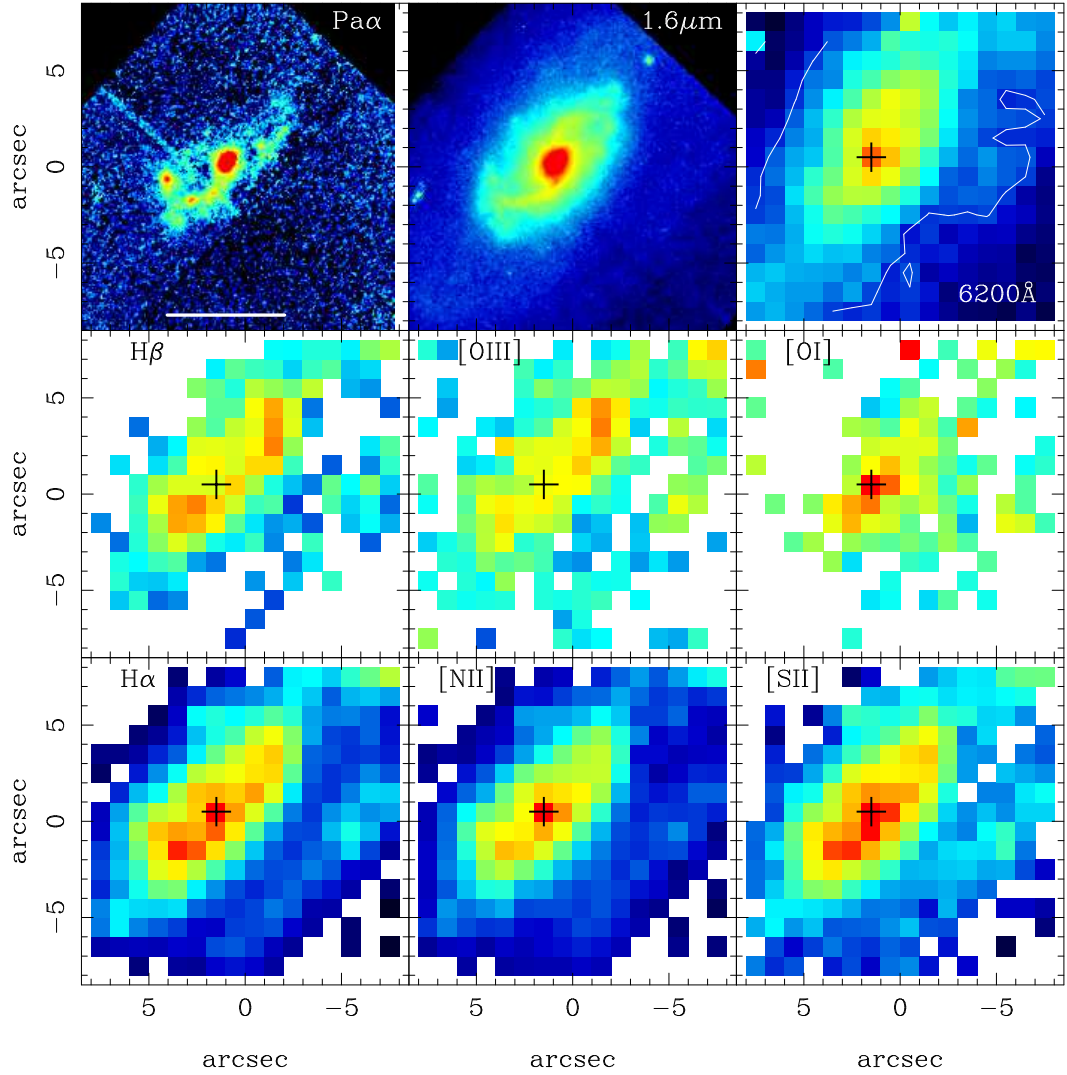}
\caption{(e) As Fig.~1a but for MCG~+02-20-003.}
\end{figure*}

\begin{figure*}
\setcounter{figure}{0}
\includegraphics[width=14.cm]{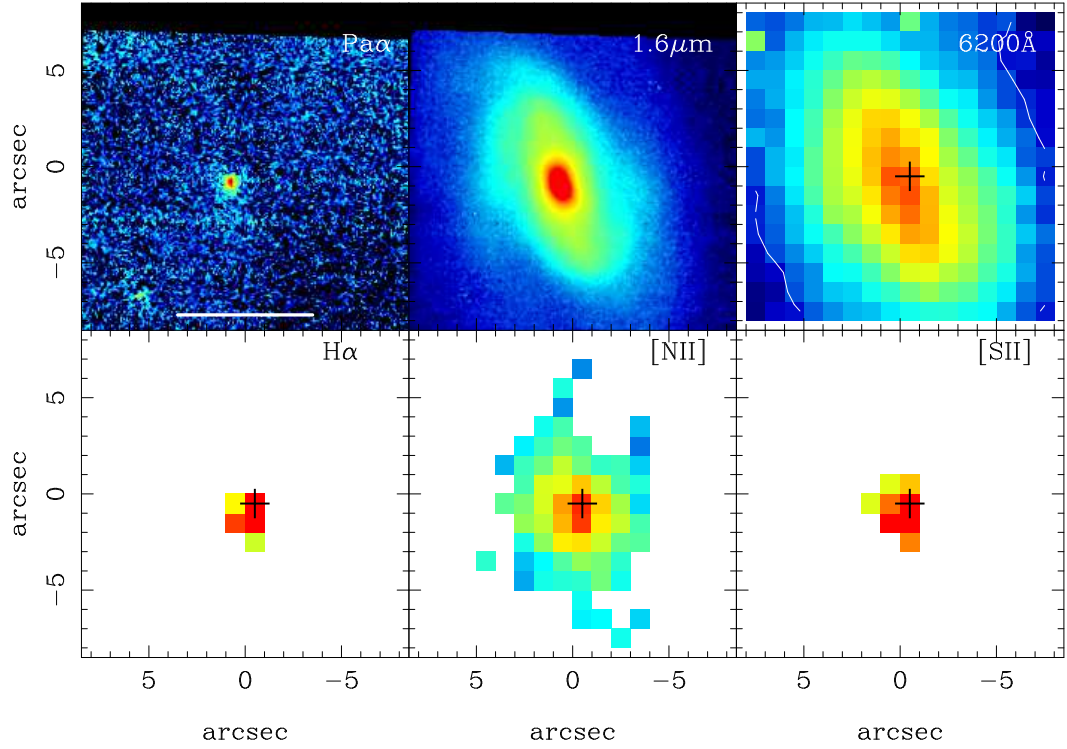}
\caption{(f) As Fig.~1a but for IC~860. The H$\beta$,
[O\,{\sc iii}]$\lambda$5007 and [O\,{\sc i}]$\lambda$6300 emission
lines are not   
detected in this galaxy.}
\end{figure*} 

\begin{figure*}
\setcounter{figure}{0}
\includegraphics[width=14.cm]{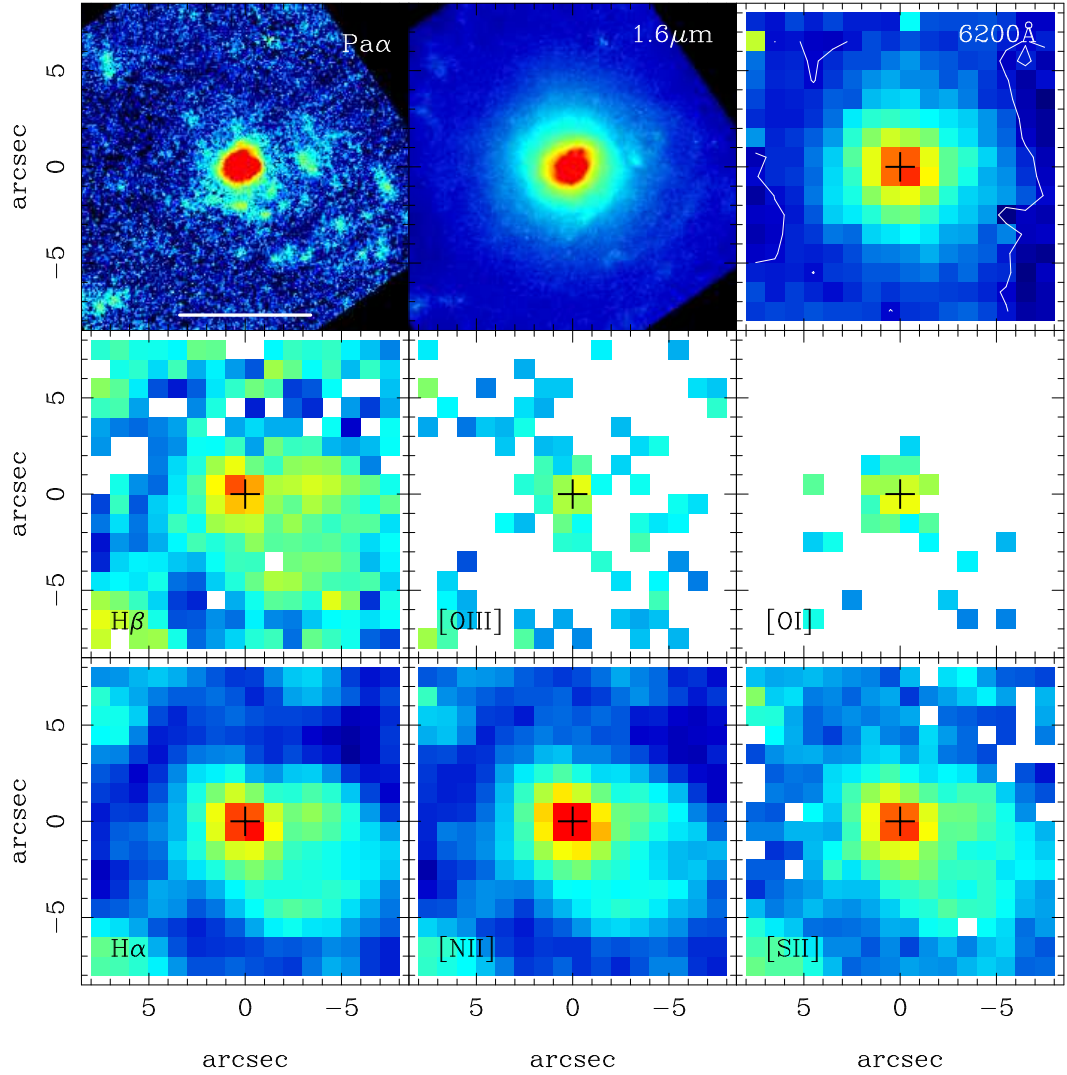}
\caption{(g) As Fig.~1a but for NGC~5936.}
\end{figure*} 

\begin{figure*}
\setcounter{figure}{0}
\includegraphics[width=14.cm]{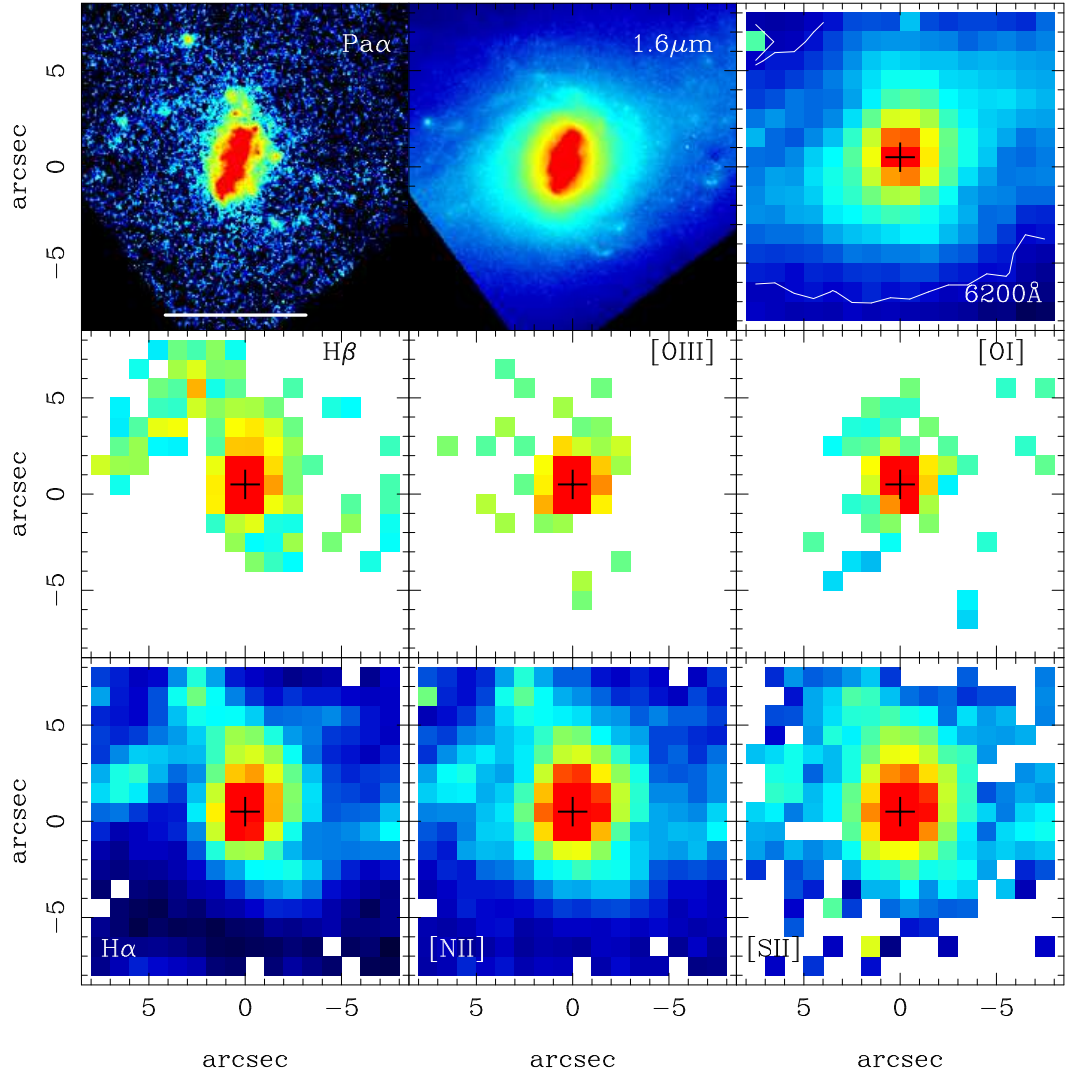}
\caption{(h) As Fig.~1a but for NGC~6701.}
\end{figure*} 

\begin{figure*}
\setcounter{figure}{0}
\includegraphics[width=14.cm]{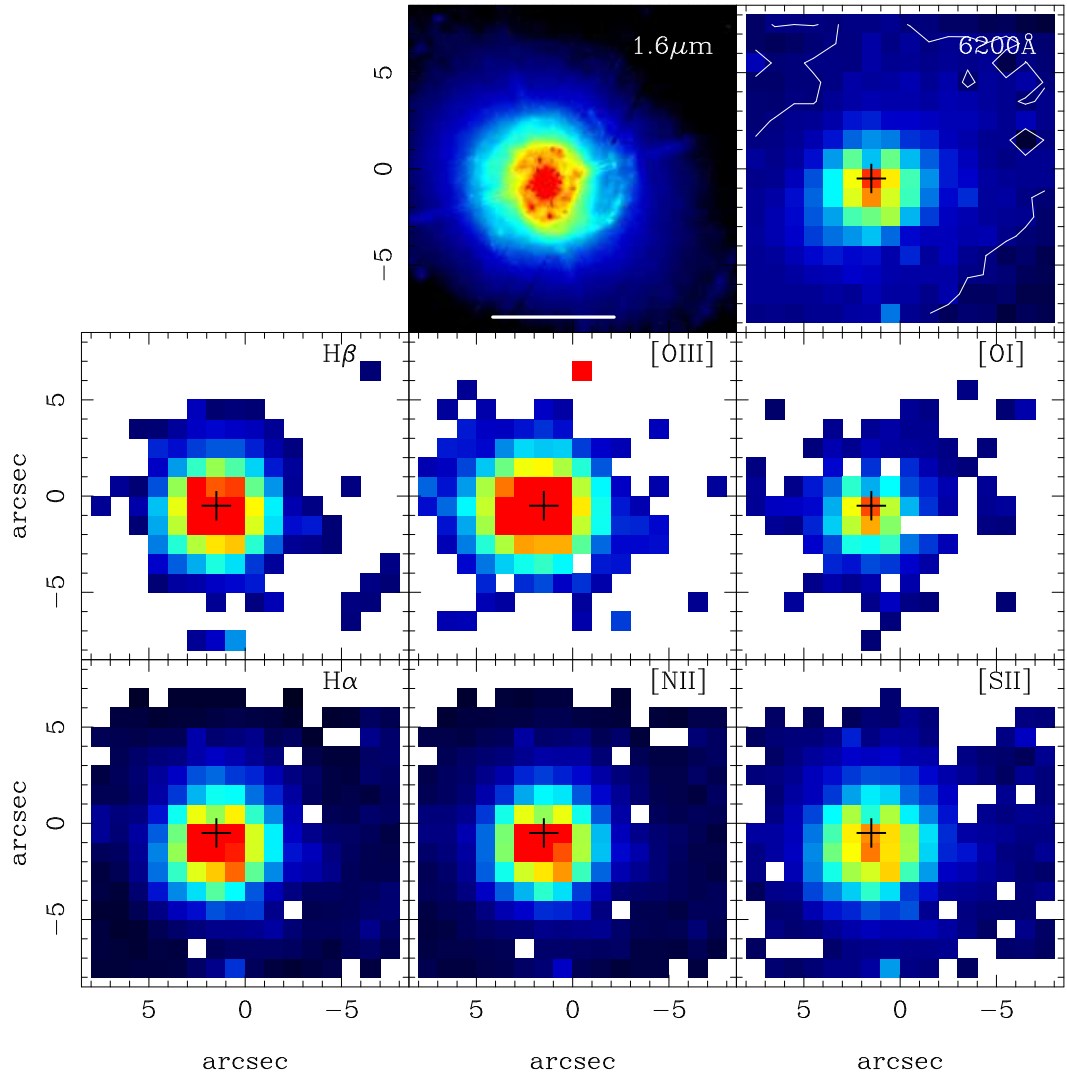}
\caption{(i) As Fig.~1a but for NGC~7469; there is no {\it HST}/NICMOS 
NIC2 Pa$\alpha$ image available for this galaxy.  The PMAS H$\beta$ and 
H$\alpha$ maps shown 
in this figure were constructed by fitting one component to the lines 
(see text and also Fig.~2).}
\end{figure*}

\begin{figure*}
\setcounter{figure}{0}
\includegraphics[width=14.cm]{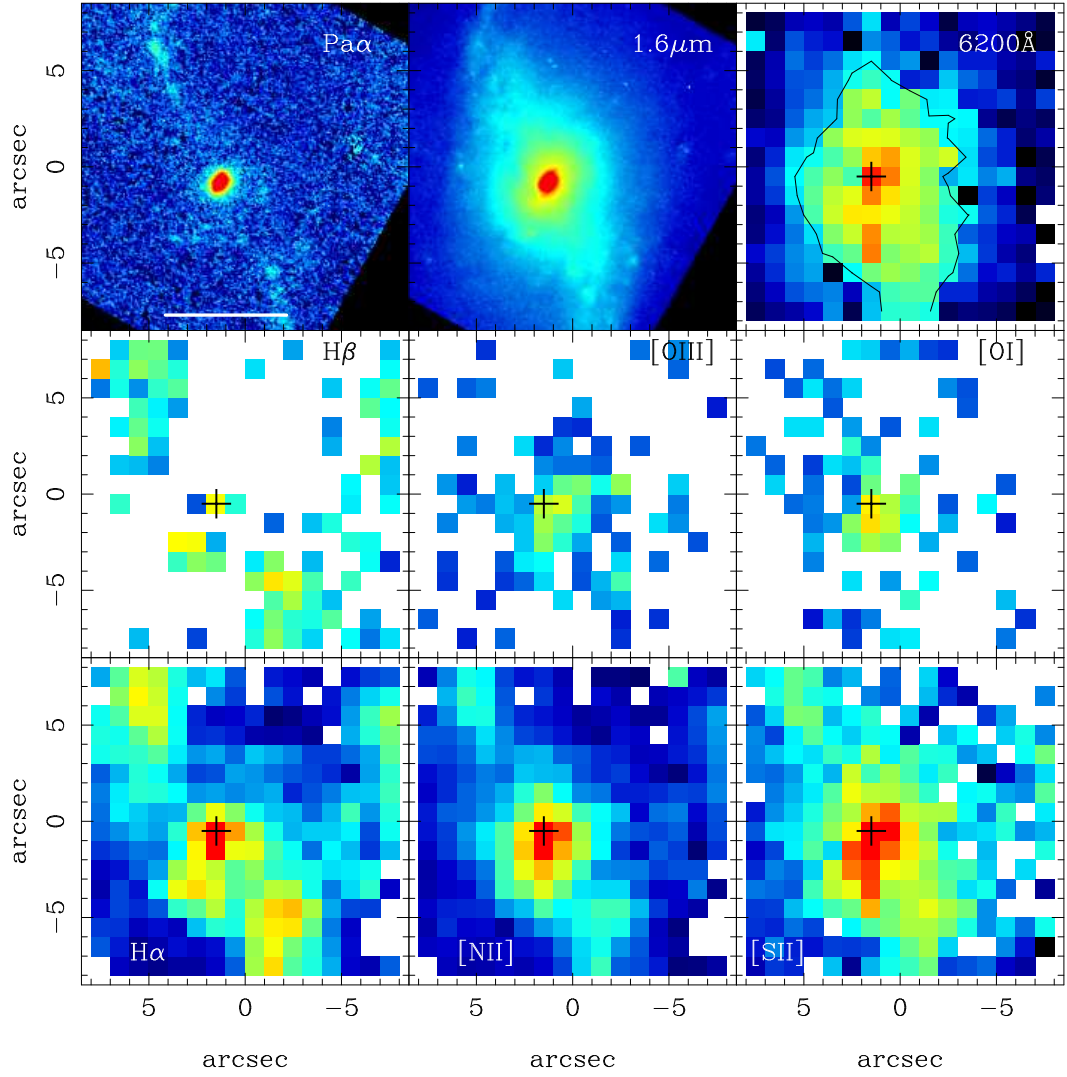}
\caption{(j) As Fig.~1a but for NGC~7591.}
\end{figure*}

\begin{figure*}
\setcounter{figure}{0}
\includegraphics[width=12.cm,angle=-90]{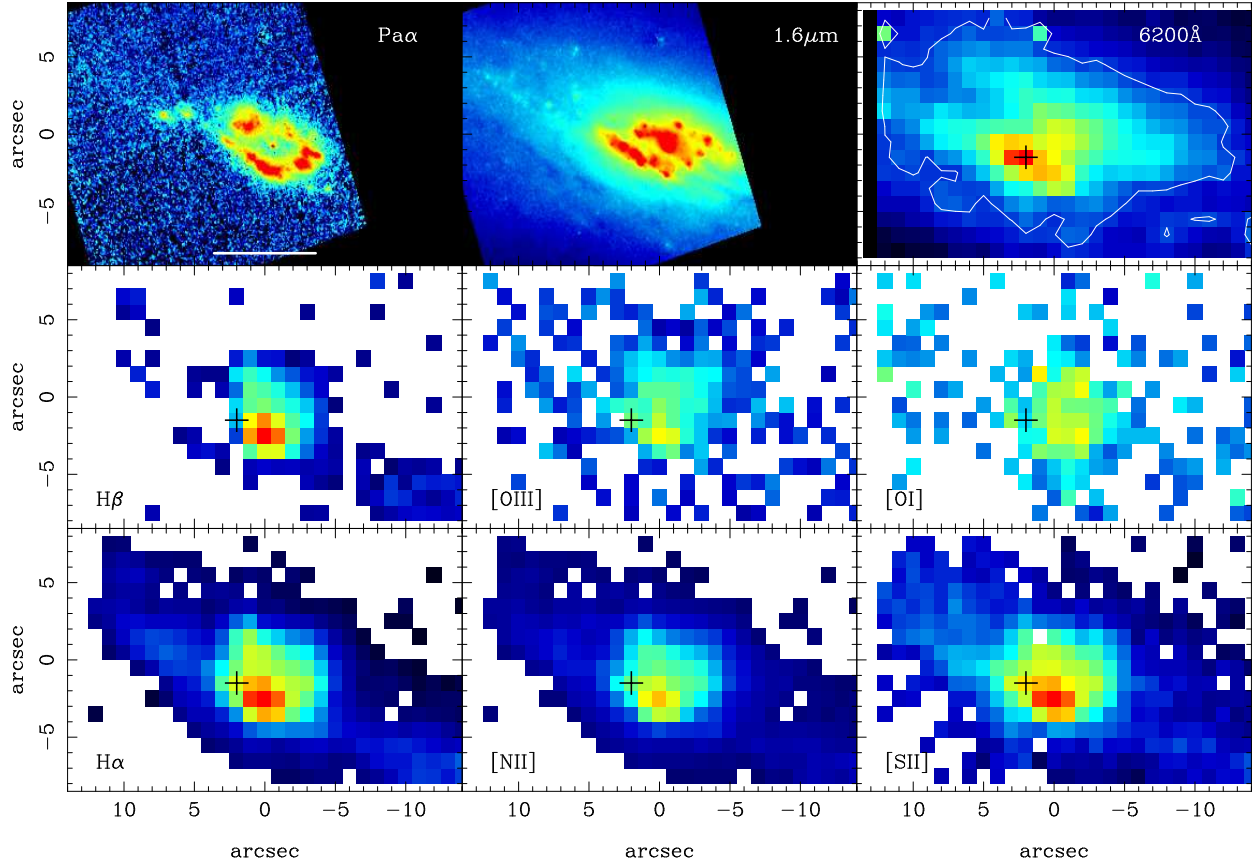}
\caption{(k) As Fig.~1a but for NGC~7771. The PMAS mosaics were constructed with
the east and west pointings done for this galaxy and they cover 
approximately the central $28\arcsec \times 16\arcsec$ region.}
\end{figure*}

\end{document}